\documentclass{aa}  

\usepackage[T1]{fontenc}
\usepackage{ae,aecompl}
\usepackage{lscape}
\usepackage[]{natbib}
\bibpunct{(}{)}{;}{a}{}{,} 

\usepackage{graphicx}   
\usepackage{amsmath}    
\usepackage{amssymb}    
\usepackage{xcolor}
\usepackage{txfonts}
\usepackage{hyperref}

\begin{document} 

   \titlerunning{DR16 SPIDERS point sources}
   \title{The final SDSS-IV/SPIDERS X-ray point source spectroscopic catalogue}
   \subtitle{}
   
\author{
J. Comparat,\inst{1}\thanks{E-mail: comparat@mpe.mpg.de}\thanks{Catalogues are available in electronic form at the CDS via anonymous ftp to cdsarc.u-strasbg.fr (130.79.128.5) or via \url{http://cdsweb.u-strasbg.fr/cgi-bin/qcat?J/A+A/}}
\and A. Merloni\inst{1} 
\and T. Dwelly\inst{1}
\and M. Salvato\inst{1}
\and A. Schwope\inst{2}
\and D. Coffey\inst{1}
\and J. Wolf\inst{1}
\and R. Arcodia\inst{1}
\and T. Liu\inst{1}
\and J. Buchner\inst{1}
\and K. Nandra\inst{1}
\and A. Georgakakis\inst{3}
\and N. Clerc\inst{4}
\and M. Brusa\inst{5,6}
\and J. R. Brownstein\inst{7}
\and D. P. Schneider\inst{8,9}
\and K. Pan\inst{10}
\and D Bizyaev\inst{10,11}
}

 \institute{
 Max-Planck-Institut f\"{u}r extraterrestrische Physik (MPE), Giessenbachstrasse 1, D-85748 Garching bei M\"unchen, Germany
\and Leibniz-Institut f\"{u}r Astrophysik Potsdam (AIP), An der Sternwarte 16, 14482 Potsdam, Germany
\and Institute for Astronomy \& Astrophysics, National Observatory of Athens, V. Paulou \& I. Metaxa, 11532, Greece
\and IRAP, Universit\'e de Toulouse, CNRS, UPS, CNES, Toulouse, France
\and Department of Physics and Astronomy (DIFA), University of Bologna, Via Gobetti 93/2, I-40129 Bologna, Italy
\and INAF-Osservatorio di Astrofisica e Scienza dello Spazio di Bologna, via Gobetti 93/3, 40129, Bologna, Italy
\and Department of Physics and Astronomy, University of Utah, 115 S. 1400 E., Salt Lake City, UT 84112, USA
\and Department of Astronomy and Astrophysics, The Pennsylvania State University, University Park, PA 16802, USA
\and Institute for Gravitation and the Cosmos, The Pennsylvania State University, University Park, PA 16802, USA
\and Apache Point Observatory and New Mexico State University, P.O. Box 59, Sunspot, NM, 88349-0059, USA
\and Sternberg Astronomical Institute, Moscow State University, Moscow, Russia
}
  
\date{Submitted, December 2019}
\abstract
{}
{We look to provide a detailed description of the SPectroscopic IDentification of ERosita Sources (SPIDERS) survey, an SDSS-IV programme aimed at obtaining spectroscopic classification and redshift measurements for complete samples of sufficiently bright X-ray sources. }
{We describe the SPIDERS X-ray Point Source Spectroscopic Catalogue, considering its store of 11,092 observed spectra drawn from a parent sample of 14,759 ROSAT and XMM sources over an area of 5,129 deg$^2$ covered in SDSS-IV by the eBOSS survey. }
{This programme represents the largest systematic spectroscopic observation of an X-ray selected sample. A total of 10,970 (98.9\%) of the observed objects are classified and 10,849 (97.8\%) have secure redshifts. 
The majority of the spectra (10,070 objects) are active galactic nuclei (AGN), 522 are cluster galaxies, and 294 are stars. } 
{The observed AGN redshift distribution is in good agreement with simulations based on empirical models for AGN activation and duty cycle.  
Forming composite spectra of type 1 AGN as a function of the mass and accretion rate of their black holes reveals systematic differences in the H-beta emission line profiles. This study paves the way for systematic spectroscopic observations of sources that are potentially to be discovered in the upcoming eROSITA survey over a large section of the sky.}

\maketitle
\section{Introduction}\label{sec:Intro}

\begin{table*}
    \centering
    \caption{\label{tab:Xagn:census}Subset of existing samples of X-ray selected AGN with spectroscopic redshift. 
    The area covered is given in square degrees. 
    The X-ray band signifies whether the sample was built using soft X-rays, hard X-rays or both. 
    FX$_{lim}$ gives the range in which the flux limits of the samples are located. 
    This is an order of magnitude, please refer to the articles to derive exact values. 
    References are 
    M05 \citet{Murray2005}, 
    S09 \citet{Salvato2009},
    B10 \citet{Brusa2010},
    S11 \citet{Salvato11},
    F12 \citet{Fotopoulou2012},
    K12 \citet{Kochanek2012}, 
    H14 \citet{Hsu14},
    N15 \citet{Nandra2015},
    Ma16 \citet{Marchesi2016}, 
    Me16 \citet{Menzel16}, 
    X16 \citet{Xue2016},
    A17 \citet{Ananna17},
    G17 \citet{Georgakakis17}, 
    L17 \citet{Luo2017},
    H18 \citet{Hasinger2018},
    L19 \citet{Lamassa2019}.
    }
    \begin{tabular}{c rr cccccc}
    \hline \hline
name & N & area    & \multicolumn{2}{c}{X-ray} & references \\
     &   & deg$^2$ & band & FX$_{lim}$ &   &  \\
\hline
SPIDERS    & 10,849 & 5128.9  & soft & [-12.5,-12]   & this paper \\
XMM-XXL-N    & 2,578 & 18.0 & both & [-15,-14] &  B10, G17, Me16\\
Stripe 82X & 1,886 & 31.3 & both & [-15,-14]  & A17, L19 \\
X-Bootes   & 2,424 & 7.7  & both & [-15,-14] & M05,K12 \\
COSMOS     & 2,169 & 2.2  & both & [-16,-15]  &  B10, S09, S11, Ma16, H18\\
AEGIS X    & 354 & 0.3  & both & [-17,-16]  &  N15 \\
CDFS       & 653 & 0.2  & both & [-17,-16] &  H14, L17 \\
CDFN       & 351 & 0.2  & both & [-16.5,-15.5] &  X16  \\
LH         & 115 & 0.2  & both & [-16,-15] &  F12 \\
\hline
    \end{tabular}
\end{table*}

\begin{table*}
    \centering
    \caption{\label{tab:catalogue:location}
    Catalogues of spectra and their links presented below. 
    The unique combination of the values in the columns \texttt{PLATE\_BEST}, \texttt{MJD\_BEST}, \texttt{FIBERID\_BEST} allows users to retrieve the corresponding spectra via the SDSS search interface. 
    The user can upload a list of identifier to retrieve the corresponding set of spectra. 
    }
    \begin{tabular}{c c}
    \hline
    \multicolumn{2}{c}{Official SDSS-DR16 Value Added Catalogues} \\
    2RXS & \object{\url{https://data.sdss.org/sas/dr16/eboss/spiders/analysis/VAC_SPIDERS_2RXS_DR16.fits}} \\
    XMMSL2 & \object{\url{https://data.sdss.org/sas/dr16/eboss/spiders/analysis/VAC_SPIDERS_XMMSL2_DR16.fits}} \\
    \hline 
    \multicolumn{2}{c}{Official data model, description of the columns} \\
    2RXS & \url{https://data.sdss.org/datamodel/files/SPIDERS_ANALYSIS/VAC_spiders_2RXS_DR16.html} \\
    XMMSL2 & \url{https://data.sdss.org/datamodel/files/SPIDERS_ANALYSIS/VAC_spiders_XMMSL2_DR16.html} \\
    \hline
    \multicolumn{2}{c}{SDSS DR16 optical spectra} \\
    \multicolumn{2}{c}{\texttt{PLATE\_BEST}, \texttt{MJD\_BEST}, \texttt{FIBERID\_BEST} at \url{https://dr16.sdss.org/optical/spectrum/search}} \\
    \hline
    \multicolumn{2}{c}{SPIDERS project web page} \\
    \multicolumn{2}{c}{\url{http://www.mpe.mpg.de/XraySurveys/SPIDERS/}} \\
    \hline
    \end{tabular}
\end{table*}

Since the advent of powerful focusing X-ray telescopes, it has become clear that the high-energy emission provides an insightful view of the extra-galactic sky. Accreting super-massive black holes dominate the number of detected X-ray sources down to the limiting fluxes detectable in the deepest pencil beam surveys today; clusters of galaxies, on the other hand, also shine brightly in X-rays due to the presence of hot plasma reaching temperatures of millions of degrees in their potential
wells. 
X-ray surveys can, therefore, be used to provide some of the most stringent constraints on the cosmological evolution of super massive black holes \citep[see e.g.][]{Hickox2017} and of the large-scale structure itself \citep[see e.g.][]{weinberg_2013_review}.
However, optical spectroscopy is almost always needed in order to unambiguously classify X-ray sources as well as measure their distances accurately. 

Over the last decade, spectroscopic observations in the optical of X-ray selected active galactic nuclei (AGN) have increased in number by about two orders of magnitude, from hundreds to tens of thousands, when combining deep and medium-deep surveys with wide area surveys \citep{Murray2005,Salvato2009,Brusa2010,Salvato11,Fotopoulou2012,Kochanek2012,Hsu14,Nandra2015,Marchesi2016,Menzel16,Xue2016,Ananna17,Georgakakis17,Luo2017,Hasinger2018,Lamassa2019}. 
A subset\footnote{\url{http://www.mpe.mpg.de/XraySurveys}} 
of existing samples of X-ray selected AGN with spectroscopic redshift is detailed in Table \ref{tab:Xagn:census}. 

In this article, we report on the spectroscopic redshift measurement of 10,849 sources for 14,759 X-ray candidates over an area of 5,128.9 deg$^2$ using the Sloan Digital Sky Survey (SDSS) telescope and spectrograph infrastructure \citep{Gunn2006,Smee13}, which constitutes the SPectroscopic IDentfication of ERosita Sources (SPIDERS) sample.

Compared to previous samples, SPIDERS covers a different parameter space in terms of area and depth and it
is also the largest X-ray point source spectroscopic catalogue to date. 
The spectroscopic data are made public in the 16$^{th}$ release of data from the  SDSS \citep[DR16][]{Ahumada2019DR16}\footnote{\url{sdss.org}}, together with two `value added catalogues', which are also part of DR16, for ROSAT and XMM-Slew sources, respectively. 
Table \ref{tab:catalogue:location} gives the links to the catalogues and a description of each column.

The SDSS-IV single fibre optical spectroscopic programme is shared between the extended Baryon Oscillation Spectroscopic Survey (eBOSS, main programme), the SPectroscopic IDentfication of ERosita Sources survey (SPIDERS, sub-programme), and the Time-Domain Spectroscopic Survey (TDSS, sub-programme), which share the focal plane during observations.
The complete SPIDERS survey programme provides a homogeneous optical spectroscopic observations of X-ray sources both point-like and extended, paving the way towards systematic spectroscopic observations of eROSITA detections over a large portion of the sky \citep{Merloni12, Predehl16, Kollmeier17, Merloni2019}. 
The programme started well before the beginning of SRG/eROSITA operations upon completing the observation of the currently existing wide area X-ray surveys. 
In particular, SPIDERS targeted sources from the ROSAT All-Sky Survey, XMM Slew sources, and XMM-XCLASS catalogues \citep{Voges1999, Voges2000, Saxton08A,Clerc12} within the SDSS-IV footprint \citep{Dawson16,blanton17}. 

Clusters of galaxies were selected by cross-correlating faint ROSAT and XCLASS extended sources with red-galaxy excess found in SDSS imaging in the range $0.1<z<0.6$ (\citealt{Clerc16,Finoguenov2019}). 
These are the most massive and largest clusters in the X-ray sky, representing a well-defined sample that can be used as a first stepping stone for cluster cosmology experiments via a measurement of the growth of structure (Ider Chitham J. et al. submitted). 
Two companion papers (Clerc N. et al. in preparation, Kirkpatrick C. et al. in preparation) describe the observation of clusters in SPIDERS.

Active galactic nuclei were selected by cross-correlating ROSAT and XMM Slew catalogues with optical and near infra-red data \citep{Dwelly17,Salvato18a}. 
In this paper, we describe the results of the observation of point-like sources. 
More specifically, we detail the case of the active galactic nuclei detected by ROSAT. 

The structure of the paper is as follows. 
We explain the data and the procedure used to construct the catalogue in Sec.~\ref{sec:Data}. 
We describe the redshifts measured in Sec.~\ref{sec:z:measurements}. 
We discuss the specific case of stars in Sec.~\ref{sec:Star}. 
Finally, we show flavour spectral stacks of type 1 AGN in Sec.~\ref{sec:SpectralAnalysis}. 
Throughout the paper, we assume the flat $\Lambda CDM$ cosmology from \citet{Planck14}. 
Magnitudes are given in the AB system \citep{Oke1983}. 

\section{Data}\label{sec:Data} 
The original SPIDERS targeting, as documented in \citet{Dwelly17}, was based on earlier versions of the X-ray catalogues than the ones that were used to build the SPIDERS-DR16 catalogues, as the X-ray-optical cataloguing methods have evolved and improved since the time of target selection. 

Here we first (in section \ref{subsec:TS:summary}) summarise the original target selection for the SPIDERS-AGN samples (based on 1RXS and XMMSL1 catalogues) and the observational completeness of these samples by the end of the SDSS-IV/eBOSS survey. 
Then in section \ref{subsec:catalog:2rxs}, we describe in detail the steps that were carried out to build the catalogues released here based on updated X-ray catalogues (2RXS, XMMSL2).  
These sections are very technical in nature. 

\subsection{Target selection summary}
\label{subsec:TS:summary}

\citet{Dwelly17} documents how the target selection was carried out on the ROSAT (1RXS) and XMM Slew v1.6 (XMMSL1) catalogues \citep{Voges1999, Voges2000, Saxton08A}. 
The area considered for target selection was the subset of the SDSS DR13 photometry footprint \citep{Fukugita96,SDSS_DR13_2017ApJS23325A} that was considered suitable for extragalactic survey work by the BOSS team\footnote{\url{http://data.sdss3.org/sas/dr9/boss/lss/boss_survey.fits}}. 
It consists of $\sim$10,800 deg$^2$ of extra-galactic sky and contains 32,408 1RXS + 4,325 XMMSL1 X-ray sources. 
For 28,515 (1RXS) and 3,142 (XMMSL1) of these X-ray sources, a counterpart was found in the AllWISE catalogue, together with an SDSS-DR13 photometric counterpart \citep[AllWISE,][]{wright10,Cutri2013}. 
11,643 (1RXS) and 1,411 (XMMSL1) of these optical counterparts had previously been spectroscopically observed in earlier phases of the SDSS project. 
Out of the 16,872 (1RXS) + 1,731 (XMMSL1) potential targets remaining, 9,028 (1RXS) + 873 (XMMSL1) passed suitability filters and were put forward for spectroscopic observation within the main SDSS-IV/eBOSS programme. 
For more details on the procedure to select the targets, please refer to \citet{Dwelly17}, particularly their Figs. 8 and 13. 
The target catalogues are available here\footnote{\url{https://sas.sdss.org/sas/dr14/eboss/spiders/target/}}.

\begin{figure}
 \centering
 \includegraphics[width=\columnwidth]{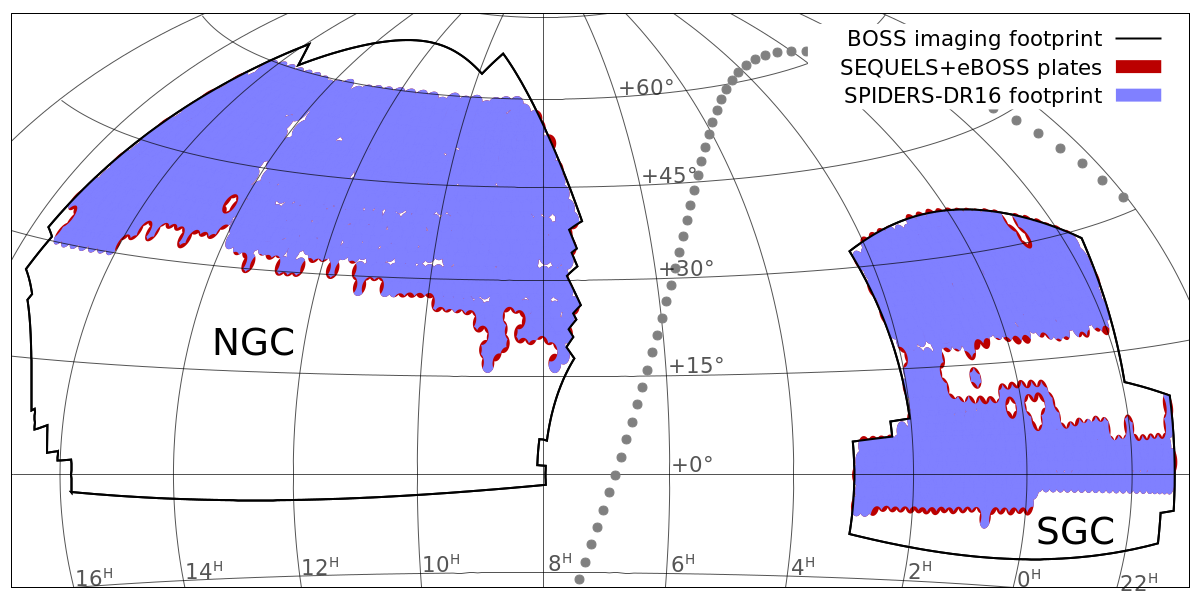}\\
 \caption{Illustration of the SPIDERS-DR16 footprint (blue, 5,129 deg$^2$) considered in this analysis, shown with an Equatorial Hammer-Aitoff projection. We also show the BOSS imaging footprint (black line, 10,800 deg$^2$), the union of all SEQUELS+eBOSS plates (red, 5,347 deg$^2$), and the Galactic Plane (grey dotted line).}
 \label{fig:mask:ra:dec:dr16}
\end{figure}

The sky area observed by the combination of the SDSS-IV/eBOSS main spectroscopic programme, plus the SDSS-III/SEQUELS pilot area, covers approximately half of the wider 10,800 deg$^2$ BOSS imaging footprint considered for the SPIDERS-AGN target selection \citep{Dawson16}. For the purposes of this paper, we define the following `SPIDERS-DR16' footprint. First we consider the sky area covered by the union of 1006 SDSS-IV/eBOSS and SDSS-III/SEQUELS plates (each plate covers a 1.49\,deg radius circle). In order to maximise the contiguity of the footprint, we  included 15 plates that do not meet the nominal eBOSS minimum signal-to-noise ratio (S/N). 
We then reject any sky areas that lie outside the BOSS imaging footprint or those that are overlapped by any plates that were planned but not observed by the conclusion of SDSS-IV/eBOSS (217.8\,deg$^2$ is rejected). 
The total remaining unique sky area in the SPIDERS-DR16 footprint is 5128.9\,deg$^2$.
Figure \ref{fig:mask:ra:dec:dr16} illustrates the SPIDERS DR16 footprints. 
Within the SPIDERS-DR16 area, there are 4,713 (1RXS) + 457 (XMMSL1) potential targets available.
We note that during the SDSS-IV observations, the focal plane was shared between three programmes: eBOSS, TDSS, and SPIDERS \citep{Dawson16,blanton17} and so there was competition for fibre resources. 
A total of 4,406 (1RXS, 93\%) + 430 (XMMSL1, 94\%) of the targets were eventually observed during the SDSS-III/SEQUELS and SDSS-IV/eBOSS campaigns. 

\subsection{The SPIDERS 2RXS sample}
\label{subsec:catalog:2rxs}

The DR16 SPIDERS 2RXS catalogue is constructed as follows. 
We consider the updated ROSAT point-source catalogue \citep[2RXS][]{Boller16} and its counterparts found via the \textsc{NWAY} software \citep{Salvato18a}. This parent catalogue does not correspond exactly to the parent catalogue used (1RXS) at the moment of targeting by \citet{Dwelly17}. 
At the bright end, higher detection likelihood, the catalogues are the same. At the faint end, marking the lower detection likelihood, there are differences. 
For a quantitative comparison between 1RXS and 2RXS, please refer to \citet{Boller16}

The 2RXS catalogue contains 132,254 sources over the entire sky, of which 21,288 lie in the SPIDERS-DR16 footprint. 

We filter the complete source list with the eBOSS footprint mask (and with a galactic latitude cut $|g_{lat}|>15^\circ$).  
We match AllWISE positions (columns names in the SPIDERS catalogue: \texttt{ALLW\_RA}, \texttt{ALLW\_DEC}) to SDSS-DR13-photo optical catalogues choosing the brightest counterpart (in \texttt{modelMag\_r}) lying within 3 arc seconds radius (larger than the 1.5 arc seconds radius used for targeting). 
In the catalogue, we select only the most likely counterpart detected in SDSS photometry as follows :
\begin{equation}
\label{eqn:A}
A = {\tt ( NWAY\_match\_flag ==1) \; \& \; (FLAG\_SDSSv5b\_best==1) }
.\end{equation}
After this, only one catalogue entry per X-ray source remains; we note, however, that in some rare cases, this is the incorrect counterpart (for example, if the uncertainty on the X-ray position is underestimated, we may miss the true counterpart if it is located beyond the search radius). 
We discuss these few cases later in the article. 
In the SPIDERS-DR16 footprint, we obtain 19,821 (10,039) X-ray sources with existence likelihoods greater than 6.5 (10).\footnote{As discussed in \cite{Boller16}, above the lower existence likelihood threshold a significant fraction (up to 30\%) of spurious sources is expected, while only about 7\% cent spurious sources are expected above existence likelihood of 10.}
We refer to these as `All' the sources of interest (labelled `A' in Figures and Tables, Eq. \ref{eqn:A}). 

Among `A', 13,986 (6,853) are in the  magnitude range to be observed by the SDSS-IV programme. 
We refer to these as candidate `targets' for spectroscopic observation with SDSS (`T', Eq. \ref{eqn:T}). 
\begin{align}
\label{eqn:T}
 T & = {\tt (SDSS\_FIBER2MAG\_i>=17) } \; \& \;  \nonumber \\
   &   {\tt (SDSS\_FIBER2MAG\_i<=22.5) } \; \& \;  \nonumber \\
   &   {\tt (SDSS\_MODELMAG\_i>=16).   }     
\end{align}
Then the SDSS spectroscopic information is added based on the optical position (using a 1.5 arc seconds matching radius between the optical source position ({\tt SDSS\_RA, SDSS\_DEC}) and fibre position on the sky ({\tt PLUG\_RA, PLUG\_DEC}). 

Among `T', 10,590 (6,145) were spectroscopically observed during one of the SDSS editions (for these, in the catalogue, the `DR16\_MEMBER' flag is set to True). 
We refer to these as `observed' (`O', Eq. \ref{eqn:O});
\begin{equation}
\label{eqn:O}
O = (T) \; \& \; {\tt (DR16\_MEMBER==1) }    
.\end{equation}

Among `O', 10,474 (6,096) were identified or classified. 
We refer to these as `identified sources' (`I', Eq. \ref{eqn:I}). 
\begin{equation}
\label{eqn:I}
I =  (c_2) | (c_3) | (c_4) | (c_5) | (c_6)                    
,\end{equation}
where 
\begin{align}
c_1 & = (O) \; \& \; (\; {\tt (Z\_BEST>-0.5) }\; |      \nonumber \\
   &  {\tt (\; (DR16\_Z>-0.5) \; \& \; (DR16\_Z\_ERR>0) \; ) \; ); }    
\end{align}
\begin{equation}
c_2 = (c1) \; \& \; {\tt (CONF\_BEST==3) }                      
;\end{equation}
\begin{align}
c_3 & = (c1) \; \& \; {\tt (CONF\_BEST==2)} \; \& \;      \nonumber \\
   &  {\tt (\; (CLASS\_BEST==``BLAZAR'')}\; | \nonumber \\
   &  {\tt (CLASS\_BEST==``BLLAC'')}\; ); 
\end{align}
\begin{align}
c_4 & = (c1) \; \& \; {\tt (DR16\_SN\_MEDIAN\_ALL>=2)} \; \& \;              \nonumber \\
   &  {\tt (DR16\_ZWARNING==0 ); }                           
\end{align}
\begin{align}
c_5 & = (c1) \; \& \; {\tt(CONF\_BEST==2) \; \& \; (DR16\_ZWARNING==0 )} \; \& \;  \nonumber \\
   &   {\tt(VI\_REINSPECT\_FLAG == 0) \; \& \; (VI\_NINSPECTORS>2);  } 
\end{align}
\begin{align}
c_6 & = (c1) \; \& \; {\tt(CONF\_BEST==2) \; \& \; (DR16\_ZWARNING==0 ) }\; \& \;    \nonumber \\
   &   {\tt(VI\_AM\_RECONCILED==1);  }                            
\end{align}

Among `I', we measured 10,366 (6,007) reliable redshifts, confirmed by visual inspection. 
We refer to these as `good redshifts' (`Z', Eq. \ref{eqn:Z}). 
The difference between I and Z consists of a set of 108 (89) featureless high signal-to-noise BLAZAR spectra, whose redshift could not be determined (classification `blazars\_noZ' below, Eq. \ref{eqn:BL});
\begin{align}
\label{eqn:BL}
{\tt blazar\_noZ } & = (I) \; \& \; {\tt (CONF\_BEST<3) }\; \& \;   \nonumber \\
   & {\tt (\; (CLASS\_BEST==``BLAZAR'')\; |}  \nonumber \\      
   & {\tt (CLASS\_BEST==``BLLAC'')\; );}        
\end{align}
\begin{equation}
\label{eqn:Z}
Z = (I) \; \& \; {\tt(blazar\_noZ == False).  }
\end{equation}

The existence likelihood, denoted {\tt exiML}, is the detection likelihood that was measured by \citet{Boller16} for the 2RXS sample ({\tt RXS\_ExiML}). 
Table \ref{tab:summary:number:sources} gives the number of object in each category A, T, O, I, {\tt blazar\_noZ}, Z for the two existence likelihood thresholds (6.5 and 10). 
The redshifts are described in detail in the following section. 

\begin{table}
    \centering
    \caption{
Number of sources in each class for the 2RXS and XMMSL2 catalogues. 
`{\tt exiML}' refers to the existence likelihood threshold applied in the X-ray. 
`Any' refers all the sources in the catalogue. For a single X-ray sources, a set of counterpart may be listed (not unique). 
`A' refers to all sources matched to their potential best optical counterpart. Each X-ray source is listed only once (Eq. \ref{eqn:A}). 
`T' refers to sources that are candidate targets for optical spectroscopic observation (Eq. \ref{eqn:T}). 
`O' refers to observed sources (Eq. \ref{eqn:O}). 
`I' refers to identified sources (Eq. \ref{eqn:I}). 
`Blazar no Z' refers to sources identified as BLAZAR for which we could not measure the redshift (Eq. \ref{eqn:BL}). 
`Z' refers to sources with good redshift measurements (Eq. \ref{eqn:Z}). 
The last column gives the targets that are uniquely present in the XMMSL2 i.e. not in the 2RXS catalogue. 
    }
    \begin{tabular}{ l r r r r}
    \hline \hline
 & \multicolumn{2}{c}{2RXS} & \multicolumn{2}{c}{XMMSL2}  \\
exiML & $>6.5$  & $>10$ & $>10$ & not 2RXS  \\ \hline

Any (non-unique)  & 21288   &   & 3196 &  \\
A. All unique     & 19821   & 10039  & 2341 & 1475 \\
T. Targets    & 13986   & 6853   & 1490 & 773  \\
O. Observed   & 10590   & 6145   & 1219 & 502  \\
I. Identified & 10474   & 6096   & 1208 & 496  \\
{\tt blazar\_noZ}   & 108     & 89     & 42   & 13   \\
Z. good Z     & 10366   & 6007   & 1166 & 483  \\ \hline
\end{tabular}
\label{tab:summary:number:sources}
\end{table}

\begin{figure}
 \centering
\includegraphics[width=0.8\columnwidth]{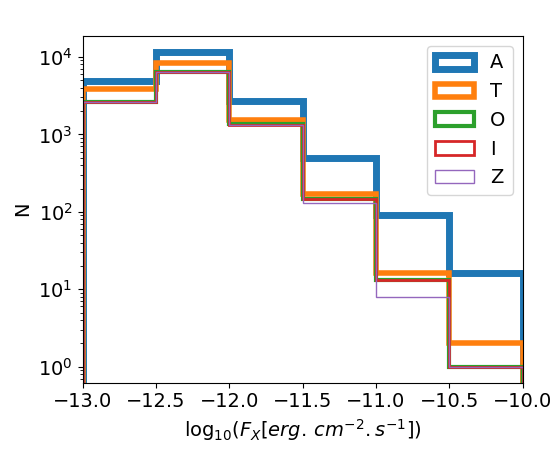}
\includegraphics[width=0.8\columnwidth]{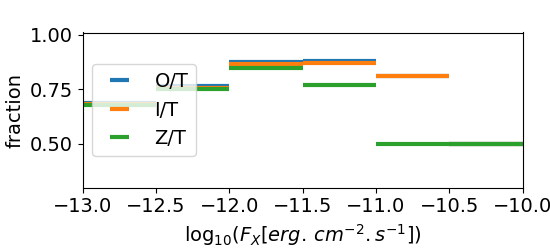}
\includegraphics[width=0.8\columnwidth]{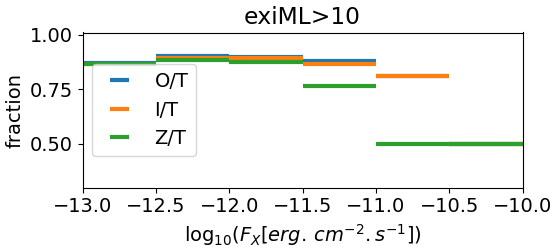}
\includegraphics[width=0.8\columnwidth]{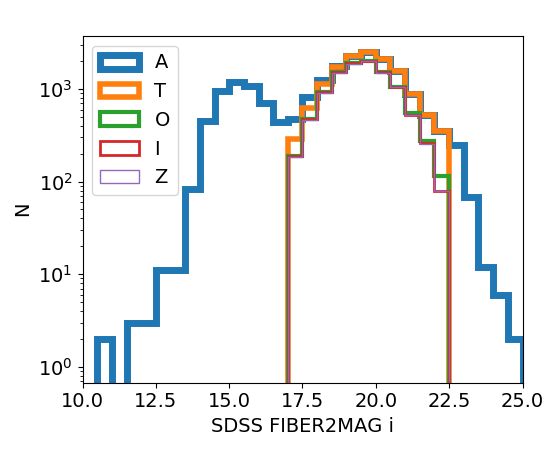}
\includegraphics[width=0.8\columnwidth]{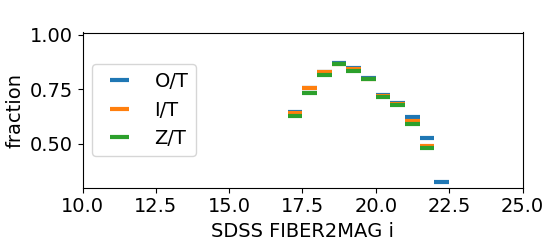}
\caption{Histograms showing the 2RXS samples with exiML$>6.5$ defined in Table \ref{tab:summary:number:sources}: A, T, O, I, Z. 
The histogram of X-ray flux shows how bright the targets are (top panel). 
The second and third panel shows the fraction of observed targets, identified objects and good redshifts with respec to the targets sample. 
The second panel is for exiML$>6.5$ and the third panel for exiML$>10$. 
They show that the exiML$>10$ Z sample is close to being a random sub sample of the targeted sample with a completeness slightly below 90\%. 
The histogram of the $i$-band fibre 2 magnitude (fourth panel) shows the impact of the optical selection made on the counterparts found, which  removes the bright objects. 
Similarly to the second panel, we show in the fifth panel the ratios O/T, I/T, Z/T as a function of fibre magnitude. 
This shows that identifying sources and determining their redshift is more difficult at the faint end.}
    \label{fig:completeness:success:rate:fluxX}
\end{figure}

We investigate the distribution of the A, T, O, I, Z samples (with exiML$>6.5$) as a function of the X-ray flux ({\tt RXS\_SRC\_FLUX}) and optical $i$-band 2 arc-seconds fibre magnitude ({\tt SDSS\_FIBER2MAG\_i}), see Fig. \ref{fig:completeness:success:rate:fluxX}. 
The X-ray flux is de-reddened from the Milky-way assuming a power law emission, which is correct for AGN but not for stars or clusters. 
The distribution of soft band X-ray flux for each sample is shown on the top panel. 
Most of the sources have a flux $-13<\log_{10}(F_X [erg.\, cm^{-2}. s^{-1}])<-11.5$. 
Few are brighter. 
The number of targets diminishes (w.r.t. all sources) as a function of flux, see curve labelled `T' (in orange).
It is due to the bright fibre magnitude and model magnitude cuts i.e. the bright X-ray sources are also bright in the optical. 
The bottom panel of the figure clearly shows the impact of the optical cuts. 
The panels showing the ratio between the observed sample and the targets as a function of X-ray flux or fibre magnitude demonstrate that the observed sample is biased with respect to the targets. Indeed the faintest and brightest objects are under-represented. 
For the high existence likelihood sample (exiML$>10$), the effect is lesser but is still present (third panel of Fig. \ref{fig:completeness:success:rate:fluxX}). 
Although we have observed 6145/6853=89\% of the exiML$>10$ targets, there remain small biases as a function of fibre magnitude and X-ray flux at the bright end. 

\subsection{The 2RXS catalogue over 10,800 deg$^2$}
\label{subsec:catalog:2rxs:fullAREA}

Over the complete BOSS extra galactic area (10,800 deg$^2$ = 2.1 times the SPIDERS-DR16 area), the total number of targets (26,685) is about twice that present in the SPIDERS-DR16 area (13,986), see Table \ref{tab:summary:number:sources:full:area}. 
Here, the fraction of observed targets is 63.1\% over 10,800 deg$^2$ instead of 75.4\% on SPIDERS-DR16,
so the completeness is lower. 
Furthermore the observed targets were chosen following different targeting schemes (previous SDSS editions), so the observed sample will be further away from being a random sampling of the complete set of targets. 
It thus complicated the statistical analysis, for example extracting an unbiased redshift distribution becomes tedious. 
This is the main reason we excluded this additional area from the catalogue and the analysis presented here. 
Using the ZWARNING=0 criterion from the SDSS pipeline (indeed inspections are not available for the complete area), we obtain an estimation of the total number of good redshifts, 16,128 (95.7\% of the observed), but cannot guarantee that all of them indeed are, due to the lack of visual inspections. 
To reach the 97.8\% of good redshifts (as in the SPIDERS-DR16 footprint) further inspection of the spectra is required. 
It would also enable the proper flagging of blazars, which redshifts are difficult to fit. 

\begin{table}
    \centering
    \caption{
Comparison of the number of 2RXS sources in each class in the SPIDERS-DR16 area and in the BOSS extra galactic area. 
    }
    \begin{tabular}{ l r r r r}
    \hline \hline
area    & SPIDERS-DR16 & BOSS \\ 
deg$^2$       & 5,129   & 10,800  \\
\hline
A. All        & 19,821   & 37,961  \\
T. Targets    & 13,986   & 26,685  \\
O. Observed   & 10,590   & 16,851  \\
Z. good Z     & 10,366   & `16,128'  \\ \hline
       
\end{tabular}
\label{tab:summary:number:sources:full:area}
\end{table}

\subsection{The SPIDERS-AGN XMMSL2 sample}
\label{subsec:catalog:xmmsl}
The DR16 SPIDERS XMMSL2 catalogue is constructed in a similar fashion to the 2RXS. 
The existence likelihood, denoted {\tt exiML}, is the maximum of the detection likelihood in any of the three bands the point source were detected in \citep{Saxton08A}\footnote{\url{https://www.cosmos.esa.int/web/xmm-newton/xmmsl2-ug}}. We have  
{\tt exiML} = Max {\tt ([XMMSL2\_DET\_ML\_B6, XMMSL2\_DET\_ML\_B7, XMMSL2\_DET\_ML\_B8]}. 
It contains 3,196 unique X-ray sources in the eBOSS footprint. 
A large fraction of them: 866, are present in the 2RXS catalogue, meaning that after removing common sources the catalogue contains 2,330 sources. 
Applying the same procedure as for 2RXS, 3,196 (2,330) sources reduce to 2,341 (1,475 not in 2RXS) sources of interest (A), 1,490 (773) targets and 1,166 (483) good redshifts, see Table \ref{tab:summary:number:sources}. 

\subsection{Summary of observations}

By combining the observations of the 2RXS and the XMMSL2 samples, we accumulated 10,849 good redshifts out of 14,759 targets over the SPIDERS-DR16 area. 
The fraction of observed targets is about O/T$\sim$73.5\% and could increase to 90-95\% with another dedicated programme. 
The fraction of identified targets among the observed is high: Z/O$=97.8\%$. 
Given that the 2RXS catalogue covers the full sky, one could extend the match to spectroscopic observation to larger areas, but the completeness would then be much lower ( O/T$\sim$30\% ) and the observed redshift may constitute a biased sample with respect to the complete sample. 
In the next decade, the combination of eROSITA with SDSS-V, 4MOST and DESI should enable the construction of a large full-sky X-ray AGN catalogue, see the discussion in Sec. \ref{sec:Discussion}.

\section{Redshift Measurement and Classification}\label{sec:z:measurements}
Automated redshift fitting for AGN is a demanding task, see \citep{Paris2012, Paris2014, Paris2017, Paris2018, Busca2018arXiv180809955B}. 
To increase confidence in the automatically obtained redshifts, we visually inspect the SPIDERS spectra. 
The visual inspection procedure and the reconciliation of results between inspectors is detailed in \citet{Dwelly17}. 
After inspection, we report the successful measure of redshifts for 97.8\% of the observed targets. 
Please refer to \citet{Menzel16} for a specific and detailed discussion on the accuracy of spectroscopic redshifts for X-ray selected AGNs. 
Overall, the number of redshift failure being quite small, we cannot study these population statistically in depth. 
Nevertheless, we see a hint that the magnitude (or fiber magnitude) distribution of undetermined redshifts is skewed towards the fainter magnitudes. 
Indeed, it should be more difficult to obtain redshift for fainter objects relative to brighter objects. 

\subsection{Classifications}

In addition to the redshift confidence flag ({\tt CONF\_BEST}), the visual inspection enable to classify in AGN types ({\tt CLASS\_BEST}). 
However, because the SPIDERS catalogues has been assembled by combining various generations of SDSS observations and visual inspections, the final classification is heterogeneous. 
For simplicity, we can group the observed objects with reliable redshift ({\tt CONF\_BEST}==3) into the following broadly defined families: AGNs (type 1 and 2), stars, AGN in clusters and galaxies in clusters. These additional classifications flags are made available here\footnote{\url{http://www.mpe.mpg.de/XraySurveys/SPIDERS/}}. 
\begin{enumerate}
    \item Stars are identified with the {\tt CLASS\_BEST==``STAR''}.
    \item Blazar: {\tt CLASS\_BEST=`BLLAC'} or {\tt `BLAZAR'}.
    \item Type 1 AGN (or Broad-line AGN, or un-obscured AGN of optical type 1), comprising {\tt CLASS\_BEST==``BALQSO'', ``QSO\_BAL'', ``QSO'', ``BLAGN''}.
    \item Type 2 AGN (or narrow-line AGN or narrow-line AGN candidates or obscured AGN of optical type 2) comprising {\tt CLASS\_BEST==``NLAGN'', ``GALAXY''}
   \item Considers the possibility of ROSAT  mistakenly identifying a source as point-like instead of extended, due to  poor PSF. 

In the latter case (5), some or all of the X-ray flux may be due to a cluster of galaxies. In order to take that eventuality into account, Galaxies or QSO are counted as possible cluster members if their redshifts are within 0.01 and their position within 1 arc minute of a redmapper cluster \citep{Rykoff2014} or a SPIDERS cluster \citep{Clerc16,Finoguenov2019}. These cannot be counted within the 2RXS or XMMSL2 X-ray flux limited AGN sample. Indeed some of the flux associated may come from the host cluster. 
\end{enumerate}

\begin{figure}
\centering
\includegraphics[width=\columnwidth]{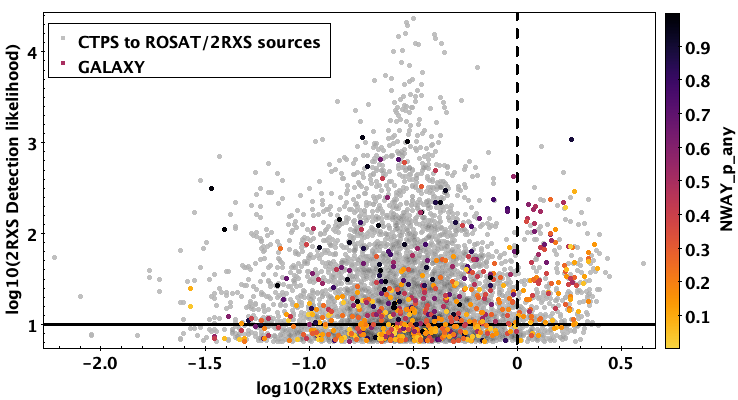}
\includegraphics[width=0.9\columnwidth]{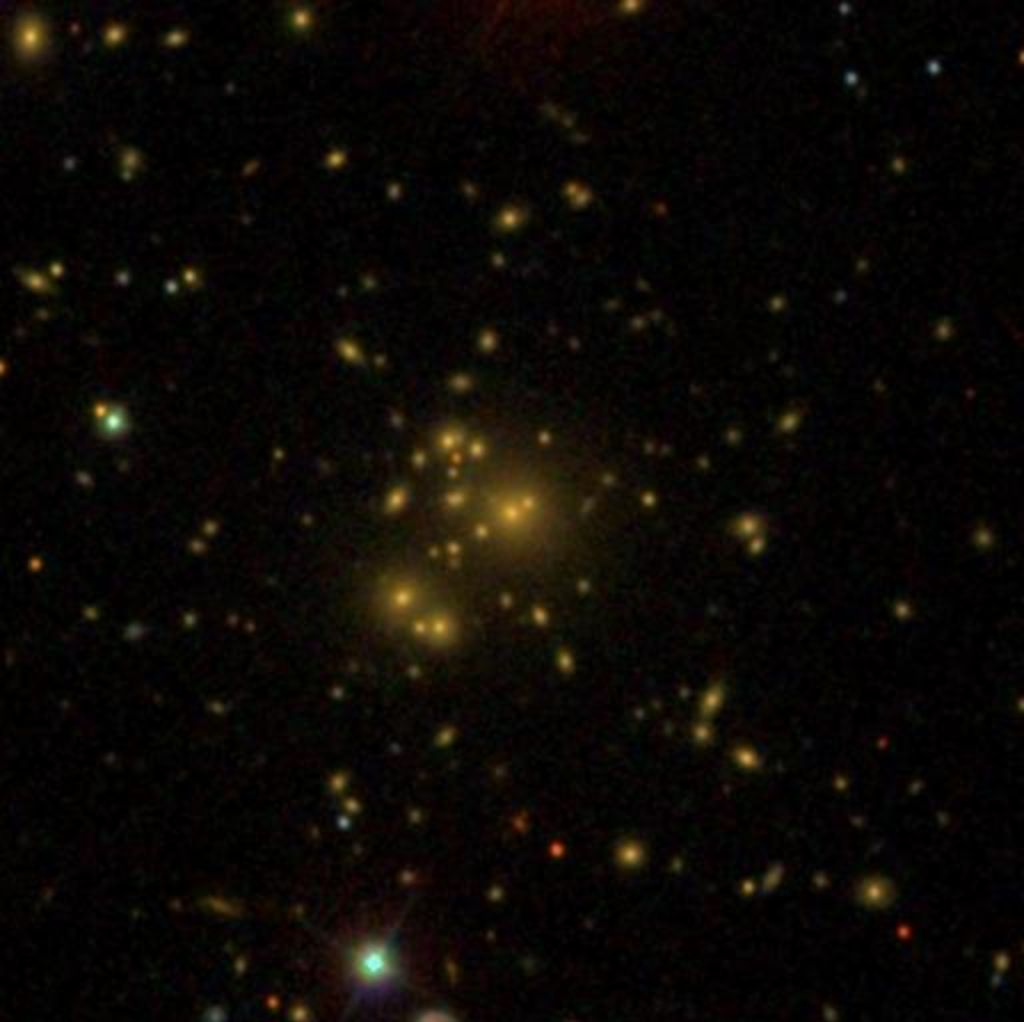}
\caption{{\bf Top}: Distribution of all sources in the X-ray detection likelihood vs X-ray extension (in arc seconds) plane. All counterparts are shown in grey (label: CTPS to ROSAT/2RXS sources). Sources classified as 'GALAXY' are coloured according to their {\tt p\_any} parameter \citep[see Sec. 3 of][for a definition of {\tt p\_any}]{Salvato18a}. The majority of the galaxies are either associated to a very low significance X-ray detection and thus just galaxies in the field (bottom part) or to extended sources (upper right part), indicating that they could be passive galaxies members of clusters or local (low redshift) extended galaxies. {\bf Bottom}: Central object in the figure is the counterpart associated to a 2RXS source, with a low {\tt p\_any} but high extension in the X-ray images. In fact, the 2RXS source in this case was extended and the associated galaxy is the central galaxy of a cluster at redshift 0.145. These type of sources populate the top/right quadrant in the top panel of the figure.}
\label{fig:ext_vs_exiML}
\end{figure}

Among the good redshift class (`Z'), after visual inspection, we list (in parentheses, separated by a plus sign) the occurrences in the 2RXS+XMMSL2 catalogue in each family: type 1 AGN (8216+941), Type 2 AGN candidates (1331+119), possible clusters members (503+62) and stars (278+27), see Table \ref{tab:redshift:classes}. 
We note that among the Cluster member candidates, 'GALAXY' refers here to the spectra without any obvious signature of an AGN. The top panel of Fig. ~\ref{fig:ext_vs_exiML} shows that these sources are usually either associated to a low X-ray source detection likelihood (and in this case the source would just be a galaxy in the field), or among the brightest members of a galaxy cluster (large extension in X-ray images, e.g., bottom panel of Fig.~\ref{fig:ext_vs_exiML}).  Most of the sources classified as 'GALAXY/Cluster' have a low {\tt p\_any} value in NWAY \citep{Salvato18a}, indicating that the reliability of the association is also low. 

\begin{figure}
 \centering
 \includegraphics[width=\columnwidth]{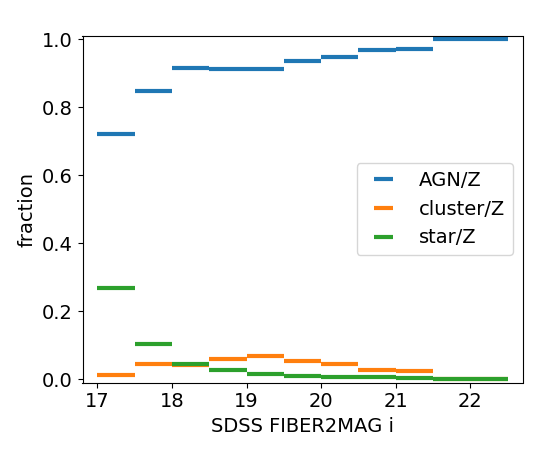}
 \caption{Fractional contribution of each object class (AGN, Cluster members, stars) to the number of good spectroscopic redshifts obtained for 2RXS sources as a function of magnitude (SDSS \tt{fiber2mag\_i})}
 \label{fig:mag:fibermag:i:histograms:classbest}
\end{figure}

We note that each population samples the fiber magnitude, model magnitude, and redshift histograms in a different fashion (see Fig. \ref{fig:mag:fibermag:i:histograms:classbest}). 
The stars sample the brighter end of the magnitude distribution. 
The AGN exclusively populate the fainter end. 
Indeed, at faint broad band magnitudes, redshift can only be determined thanks to strong emission lines; and the galaxies in clusters sample intermediate magnitudes. 

\begin{table}
    \centering
    \caption{
Number of identified redshift in each class: AGN (Type 1 or 2), Potential Cluster Members, STAR. 
`exiML' refers to the existence likelihood threshold applied in the X-ray catalogue. 
`Z' refers to sources with good redshift measurements. 
    }
    \begin{tabular}{ l r r r r}
    \hline \hline
 & \multicolumn{2}{c}{2RXS} & XMMSL2\\
exiML & $>6.5$  & $>10$ & $>10$   \\ 
Z                  & 10366        & 6007 & 1166 \\  \hline
Type 1 AGN         & 8216         & 4904 & 941 \\
Type 2 AGN         & 1331         &  602 & 119 \\
BLAZAR AGN         & 38           &   51 & 17 \\
Cluster   & 503 &  362  & 62  \\
-(GAL/QSO) & (387/116) &  (264/98) & (40/22) \\
STAR               & 278          &   88  & 27   \\ \hline
\end{tabular}
\label{tab:redshift:classes}
\end{table}

\subsubsection{AGN}

Among the AGN, the majority (8216/9622$\sim85\%$) show a spectrum with broad features \citep[emission line widths in excess of 200 km/s][]{Bolton12}. We name these type 1 AGN. 
1331 are classified as type 2 AGN. 
The remaining few are either BLAZAR or broad absorption line QSO. 

The type 2 AGN category is constituted by heavily obscured AGN (or candidates). 
Among the 1331, 602 (729) have a high (low) existence likelihood in the X-ray (i.e. above and below 10). 
For the population of high existence likelihood, the spurious fraction expected is of order of 7 per cent, that is, about 40 among 602. 
The spurious fraction should be higher (about 50\%) among the 729 with low existence likelihood. 
More accurate X-ray observations, deeper optical data, and a detailed emission line analysis are needed to disentangle these cases. 
We leave such analysis for future studies, and note that machine learning algorithm using spectral features may be a key in this process \citep[e.g.][]{Zhang2019}.  

\begin{figure*}
    \centering
    \includegraphics[width=2\columnwidth]{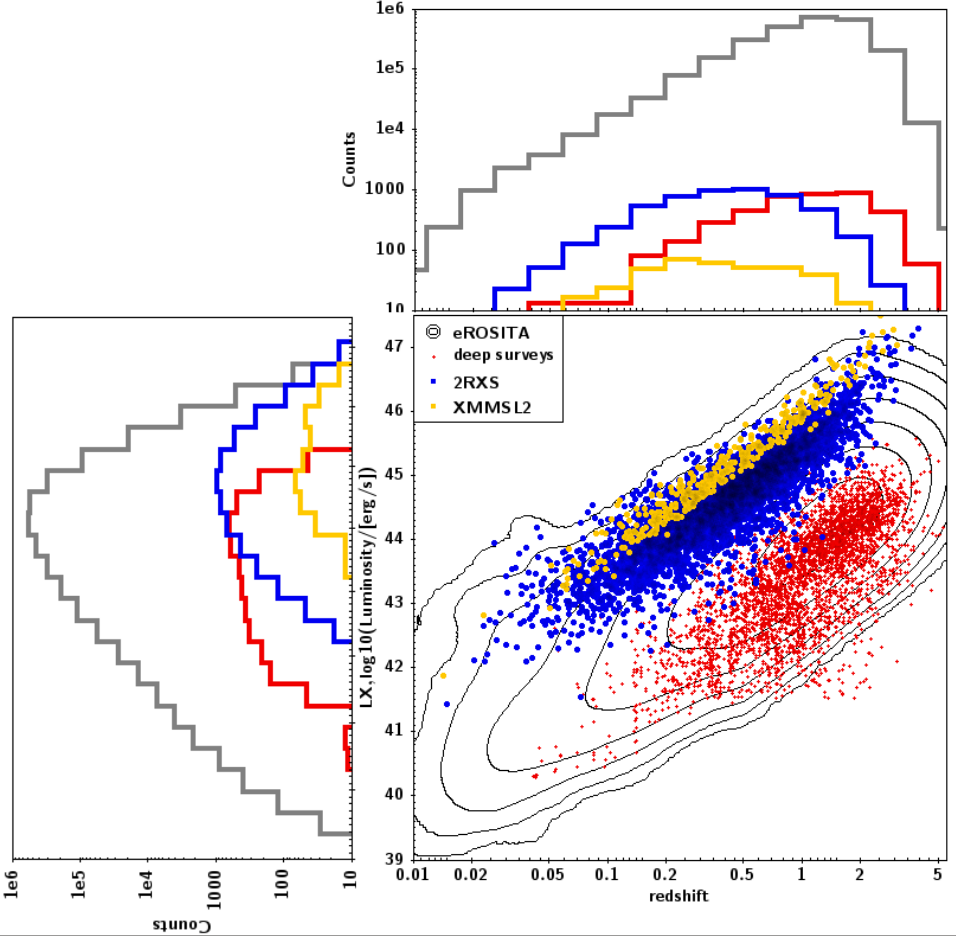}
    \caption{AGN X-ray luminosity vs. redshift for the 2RXS (blue) and XMMSL2 samples (yellow) and comparison with deep pencil beam surveys listed in Table \ref{tab:Xagn:census} (red crosses) and the prediction for the upcoming eROSITA sample (grey). The samples shown cover the plane in a complementary fashion. 2RXS (XMMSL2) X-ray luminosities are computed in the bands 0.1-2.4 (0.2-12) keV.}
    \label{fig:LX:Z}
\end{figure*}

Following Sec. 5 of \citet{Coffey2019}, we compute the 2RXS (XMMSL2) X-ray luminosities in the bands 0.1-2.4 (0.2-12) keV. 
The 2RXS (XMMSL2) flux is modelled with an absorbed power law, \texttt{mod pha*powerlaw}, with a slope of $\Gamma = $2.4 (1.7) with the n$_H$ set to that of the Milky Way.  
Figure \ref{fig:LX:Z} shows the X-ray luminosity vs. redshift for the 2RXS (XMMSL2) samples. 
It compares them to a set of the deep pencil beam surveys referenced in Table \ref{tab:Xagn:census} (red points) and the upcoming eROSITA sample (purple) taken from the mock catalogue of \citet{Comparat19}. 
The three data sets are very complementary in sampling the redshift luminosity plane. 
The SPIDERS-DR16 sample will participate to a more quantitative estimate of the evolution with redshift of the bright end of the X-ray AGN luminosity function \citep{Miyaji2000,Miyaji15,Aird15,Georgakakis17}. 
Indeed, this sample has a comparable number of sources to all pencil beam surveys together.

\subsubsection{Clusters members}

We find 503+62 sources in clusters, 374+36 (72\%) feature a galaxy  spectrum ({\tt CLASS\_BEST==``GALAXY''}) and the remaining have typical AGN spectrum. 
Overall, the contamination by galaxies in clusters is small, of order of 3\% (374/10368). 

\subsubsection{Stars}

A complete section on stars is presented in Sec. 4. 

\subsection{Redshift distribution}

\begin{figure*}
    \centering
    \includegraphics[width=2\columnwidth]{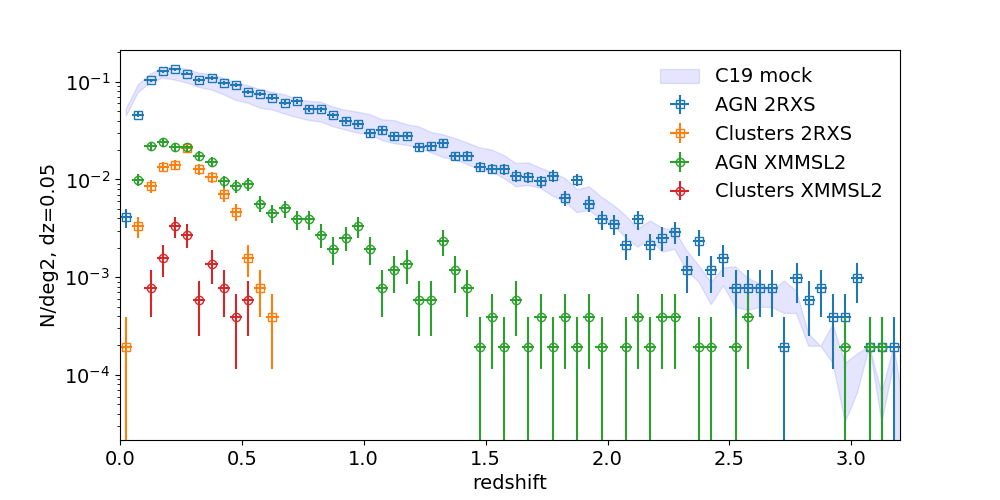}
    \caption{Observed redshift distribution of the SPIDERS-DR16 2RXS and XMMSL2 samples. 
    The C19 shaded area shows the prediction based on the mock catalogues from \citet{Comparat19}. 
     }
    \label{fig:NZ:dr16}
\end{figure*}

We find that the redshift distribution observed has the shape expected for an X-ray flux-limited sample with a broad optical magnitude range cut. In Fig. \ref{fig:NZ:dr16} 
we show the redshift distribution observed per square degrees in the 2RXS and XMMSL2 catalogues for each classification: AGN and cluster. 
For XMMSL2, which has the brightest flux limit (log10 around -12) the number density per unit sky area increases and reaches its peak in the bin $0.1<z<0.2$. 
For 2RXS, which has a fainter flux limit (log10 around -12.5) the peak in number density occurs in the bin $0.2<z<0.3$. 
It compares favourably with predictions from an adaptation of the mock catalogue of \citet{Comparat19}. 
To adapt the mock sample, we re-sample the X-ray fluxes and optical magnitudes to match the depths of the 2RXS catalogue and of the SDSS optical photometric survey. 
There is a discrepancy at low redshift: a deficit of AGNs in the observed sample compared to the mock. 
It is due to the bright magnitude and fiber magnitude cuts applied to the targeted sample; see Fig. \ref{fig:completeness:success:rate:fluxX}. 
Indeed these cuts remove a part of the low redshift AGNs, but they are difficult to mock properly.

\begin{figure}
\centering
\includegraphics[width=\columnwidth]{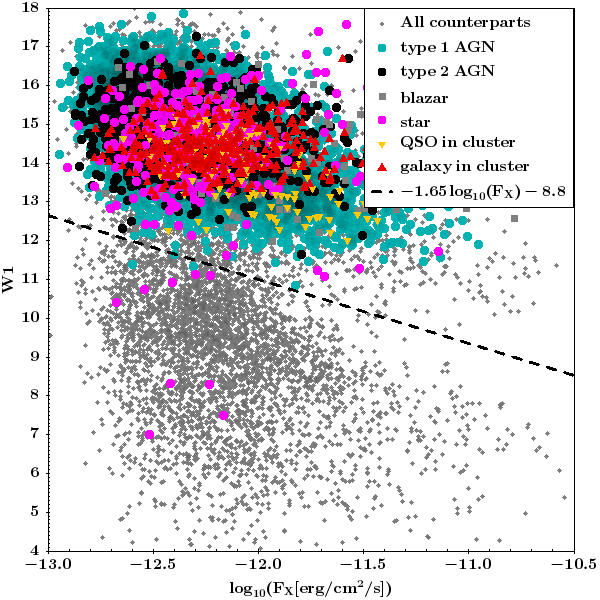}
\caption{W1 magnitude vs X-ray flux for 2RXS sources without (grey) and with (colored) spectroscopy, as labeled. The dashed line, taken from \citet{Salvato18a}, define the loci of AGN and compact objects (above) and stars (below). Note that here AGN contains both type 1 and type 2 (and candidates) objects.
}
\label{fig:wise_X_gal}
\end{figure}

We complemented the SPIDERS-DR16 catalogue with a variety of multi-wavelength information: X-ray \citep[2RXS, XMMSL2,][]{Boller16, Saxton08A}, optical \citep[SDSS,][]{SDSS_DR13_2017ApJS23325A}, \citep[GAIA,][]{Gaia2018}, infra-red \citep[AllWISE,][]{wright10}, radio \citep[FIRST,][]{White1997}. 
As in \cite{Salvato18a}, we plotted in Fig. \ref{fig:wise_X_gal} the W1 magnitude vs the X-ray flux of the sources, adopting the same line that was suggested to be able to separate AGN and compact objects from stars.

\begin{figure}
\centering
\includegraphics[width=\columnwidth]{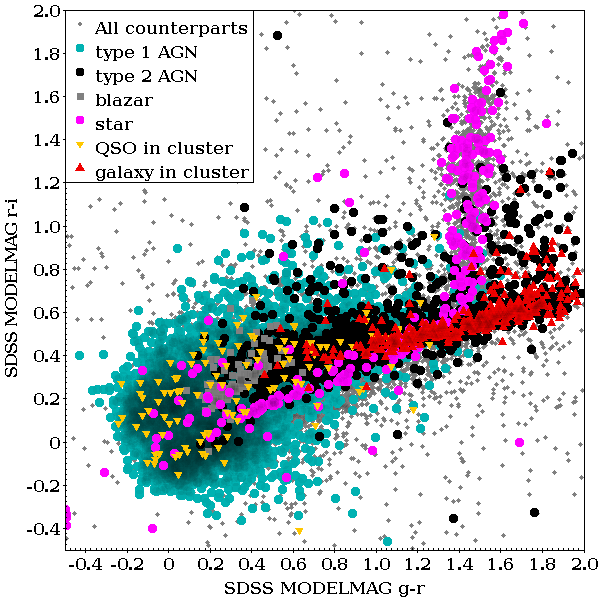}
\caption{r-i vs. g-r colors (from SDSS MODEL MAG) for the counterparts to 2RXS sources, split by their spectroscopic classification, as labeled.}
\label{fig:sdss_col}
\end{figure}

\begin{figure}
\centering
\includegraphics[width=\columnwidth]{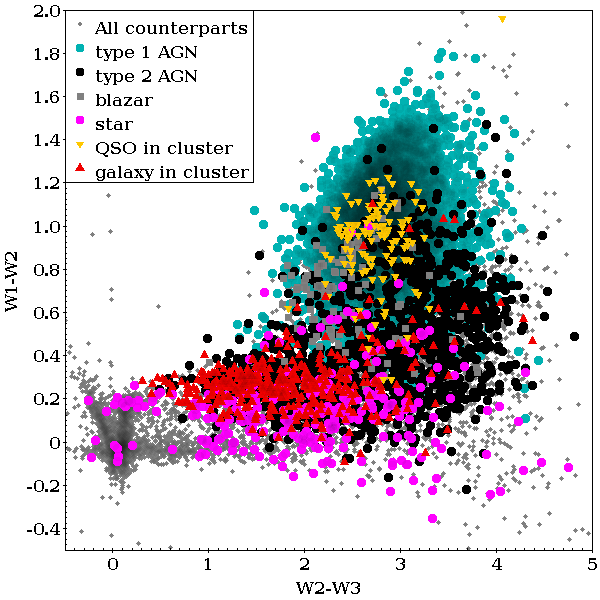}
\caption{W1-W2 vs. W2-W3 colors (from WISE) for the counterparts to 2RXS sources, split by their spectroscopic classification, as labeled.}
\label{fig:wise_col}
\end{figure}

Figure \ref{fig:sdss_col} shows the SDSS  g-r vs. r-i colours for all our SPIDERS sources. The vast majority of AGN cluster around a blue locus (g-r$<0.5$ and r-i$<0.5$) 
Sources classified as BLAZAR lie in the same blue locus.  Some AGN are redder (obscured) and thus extend to the to right corner of the plot.  
The sequence of stars also appears clearly.  
Galaxies in clusters are mostly red and QSO in clusters are mostly blue. 
A consistent picture emerges also from the analysis of the  with the WISE colour-colour diagrams (W1-W2 vs. W2-W3) shown in Fig. \ref{fig:wise_col}. 
The interplay of the different classes in color-magnitude space should open a new window to determine optimal priors to select counterparts \citep{Salvato18a}. 

\section{X-ray Stars}\label{sec:Star}
Visual screening of all spectra obtained in the SPIDERS programme and of those obtained during earlier phases of the SDSS programme and associated with 2RXS and XMMSL2 sources led to a separation of stellar objects from the large body of extra-galactic objects. The 2RXS and XMMSL2 catalogues list 290 and 37 stellar objects with attribute {\tt CLASS\_BEST==`STAR'}, but 278 and 27 only, when the criteria described in Sect.~\ref{sec:Data} are applied. The number 27 is further reduced to 16, when duplications with the 2RXS catalogue are removed. 

Obtaining a spectrum of an object classified as 'STAR' does not entail that the counterpart of the X-ray source has been identified; for this, a second X-ray identification screening step (XID) is needed. 
While the initial screening was undertaken by several individuals and a compromise had to be found in case of deviating results (classification, redshift), the XID screening step was performed by just one of the authors (AS) with the potential risk of introducing some biases or errors, but the potential advantage of a more homogeneous way of classifying stars. 
Screening for XID was done with the help of a few extra data products. These were: (a) an optical finding chart based on a PanSTARRS \citep{flewelling+16} $g-$band image (location of the X-ray centroid, the X-ray uncertainty and the target indicated), an X-ray to optical colour-colour diagram ($\log(f_{\rm X}/f_{\rm opt})$ vs.~$g-r$), and a long-term light curve obtained from the Catalina Real-Time Transient Survey \citep[CRTS, ][]{drake+09}. For almost all targets, the 'EXPLORE' feature of the SDSS-sciserver was used to search for possible other counterparts and to search for entries in the SIMBAD or NED databases.

Based on the available information, a first decision was made if the object could be confirmed as a star. 
This first screening step was performed on the more general {\tt CLASS\_BEST==``STAR''} sample and led to a revision of a number classifications that are documented in Table \ref{table:xid}. 
We corrected the incorrectly labelled source {\tt CLASS\_BEST==``STAR''} in appendix Table \ref{table:classbest:correction}. 
Then a second decision about the reliability of the target being the counterpart of the X-ray source was made. 
An XID-flag was assigned to each spectrum indicating this kind of reliability, ranging from XID=1 to XID=3.
XID=1 means that the object is regarded being the optical counterpart with high confidence. XID=2 means that the object could be the counterpart or at least could contribute to the observed X-ray emission. 
This often means that some typical ingredient or hallmark is missing or that the object seems to be blended or shows other morphological complexities.
An XID=3 object is regarded likely not being the counterpart of the X-ray source. 
 Table \ref{table:xid} in the Appendix contains the results and XID values for the objects classified as stars. 

All stellar targets were sub-classified into three main classes: coronal emitters (including flare stars), white dwarfs (WD), and compact white-dwarf binaries, either in a detached or a semi-detached configuration. The latter are the cataclysmic binaries, were a white dwarf accretes matter from a main-sequence star via Roche-lobe overflow. The break-down of stars flagged XID=1 into those three main sub-classes for the 2RXS and the XMMSL2 samples is given in Tab.~\ref{t:starsxid}.
In the star-related Tables, we use the following acronyms to classify the sources: 
\begin{itemize}
 \item CV: cataclysmic variable with unknown sub-category
 \item CV/AM: cataclysmic variable of AM Herculis type
 \item CV/DN: cataclysmic variable of dwarf nova type
 \item WDMS: detached white dwarf/main sequence binary
 \item LARP: low accretion rate polar
 \item DB+M: a binary consisting of a white dwarf of spectral type DB and a companion star
\end{itemize}

\begin{table}
\caption{\label{t:starsxid} Breakdown of 2RXS and XMMSL2 objects with high confidence identifications (XID=1) into three main object classes}
\begin{center}
\begin{tabular}{lrrr}
\hline
SUBCLASS & 2RXS & XMMSL2 & not in 2RXS\\
\hline
Coronal emitters & 61 & 3 & 3\\
WDs & 6 & -- & --\\
Compact WD binaries & 35 & 16 & 5 \\ 
\hline
\end{tabular}
\end{center}
\end{table}

\subsection{2RXS} 

\begin{figure}
 \centering
 \includegraphics[width=\columnwidth]{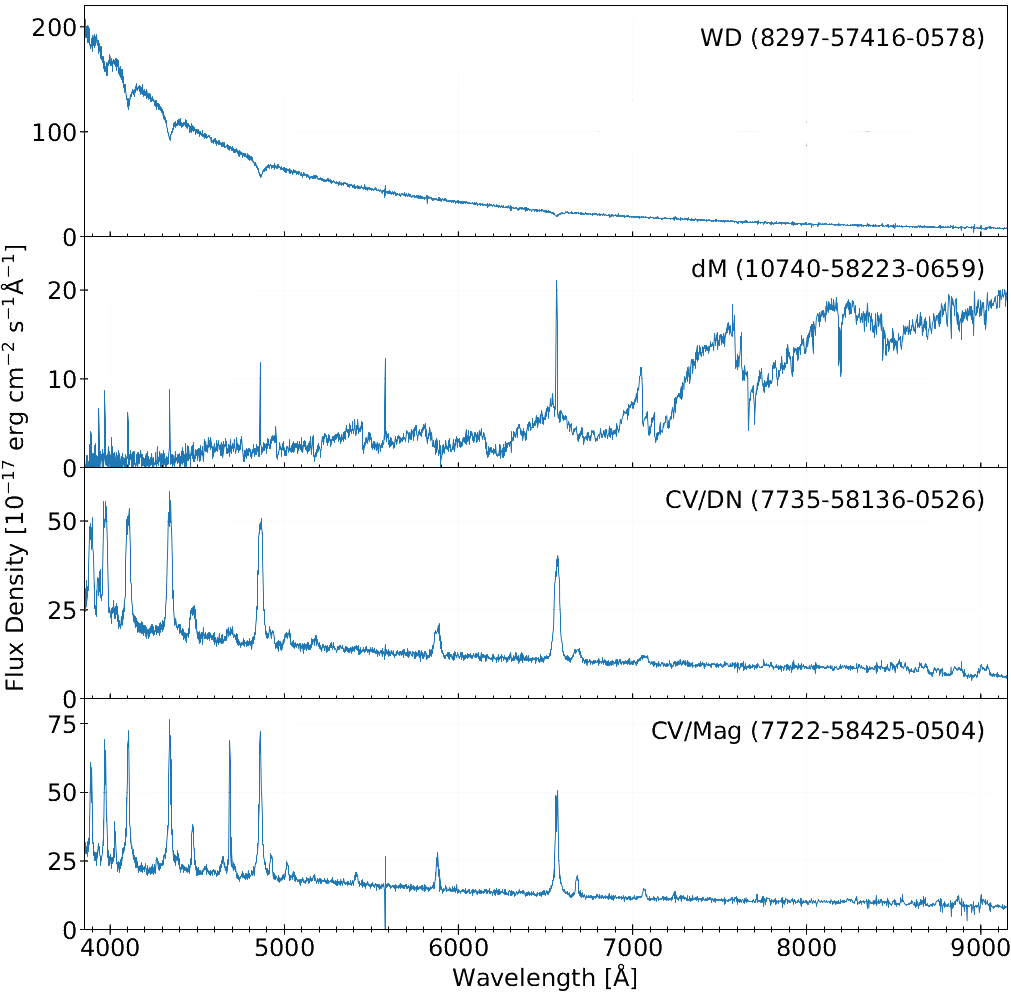}
 \caption{\label{f:stars} 
 Sample spectra of XID=1 objects, a hot white dwarf, a flare
star, a non-magnetic cataclysmic variable (dwarf nova), and a strongly
magnetic cataclysmic variable (a polar or AM Herculis star).
The PLATE-MJD-FIBERID combination and the type of X-ray emitter are
indicated in the panels. All spectra were obtained in the current SDSS programme.}
 \end{figure}

The distribution of stellar spectra over the three XID bins (1/2/3) is (102/77/99).
Among the 102 XID=1 sources from the 2RXS list, we find 67 single stars (coronal emitters and hot or sufficiently close white dwarfs) and 33 binaries with a compact object, most of them (29) being cataclysmic variables (CVs). Sample spectra of those typical X-ray emitters are displayed in Fig.~\ref{f:stars}. 

Interestingly, 75 of the 102 high-confidence (XID=1) counterparts have an NWAY ${\tt p\_any} < 0.5$, illustrating the fact that the Bayesian prior used in the X-ray to IR/optical association seems to disfavour true stellar X-ray emitters. For a stellar survey, a different prior is needed. 

\noindent We list the reasons for an XID=2 classification over an XID=1:
(1) the object appeared optically too faint for the given X-ray flux, 
(2) an M-star did not show any obvious sign of activity like H$\alpha$ in emission of flares/flickering of the light curves, 
(3) large X-ray positional errors could cast doubt on the uniqueness of the identification, in particular if the object does not show strong signs of activity which, together with an atypical optical faintness casts doubt on the reliability of the X-ray to optical association,
(4) apparent binaries were found, so that the X-ray-WISE-SDSS association chain led to ambiguities (an unresolved double WISE counterpart to the X-ray source was associated with the wrong SDSS object), 
(5) the contribution of the WISE-blended source could not be quantified.   

An example of such an of XID=2 classification is  
J002317.1+191028 (7590-56944-674), which is an M-star showing H$\alpha$ in emission and displays a variable light curve, hence qualifies as X-ray emitter, although being found with an uncomfortably large $f_{\rm X}/f_{\rm opt}$. We found that a QSO, SDSS J002319.72+190958.2, at redshift $z=1.504$ with a similar distance to the X-ray position and could contribute to the X-ray flux or even dominate. This object was thus put in the XID=2 bin because both objects could contribute to the X-ray emission.

XID=3 sources were classified as such mainly for two reasons: 
(a) the targeted object was too faint with high confidence for being compatible with a stellar coronal emitter, meaning that it had a too high an X-ray flux or a too faint an optical brightness to be compatible with the maximum $L_{\rm X}/L_{\rm bol}$ which was assumed to be $\leq -3$ 
(b) another much more typical X-ray emitter was found (often even closer) to the X-ray position (e.g. an A0 star was targeted (3454-55003-211), one of the least X-ray active stars, but a white dwarf SDSS J155108.25+454313.2 was found to lie closer to the X-ray position). 
Indeed most of the discarded objects had QSOs, CVs or WDs as more likely counterparts. 
These more likely counterparts already had spectra taken by previous editions of SDSS, so in the SPIDERS programme, they were targeted as possible secondary sources to investigate their hierarchy. 

\begin{figure}
 \centering
 \includegraphics[width=\columnwidth]{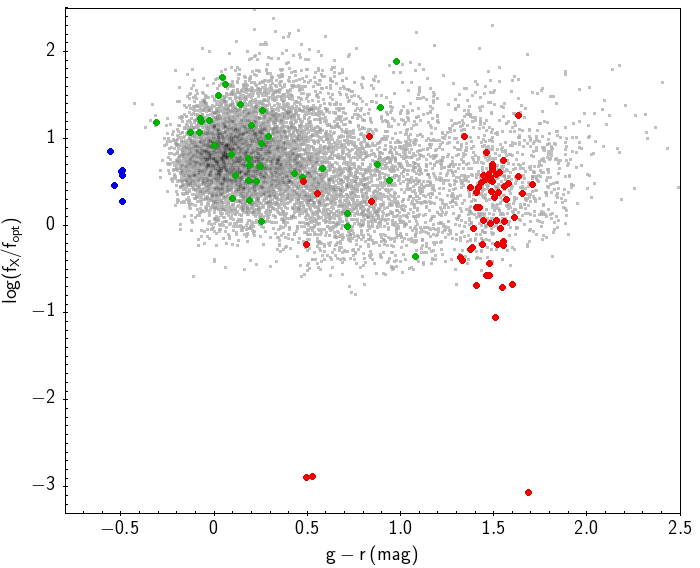}
 \caption{X-ray/optical colour-colour diagram highlighting the XID=1 objects on the background of all identified objects of the 2RXS sample. Hot white dwarfs, coronal emitters and close binaries are shown with blue, red and green symbols, respectively.}
 \label{f:stars_ccd}
 \end{figure}
 
A further two X-ray sources were associated with M-stars (spectra with PLATE-MJD-FIBERID 693-52254-0599 and 1046-52460-0078) but had unusually large X-ray positional uncertainties. 
Inspecting the area around the M-stars revealed many galaxies with concordant redshifts, obvious clusters of galaxies with the BCG rather close to the targeted star.
While the spectrum taken was clearly that of a star, the X-ray source was likely not point-like. 
While these two objects were most pronounced and for that reason discussed here separately, there are possibly more of this kind in the larger sample. As stated above, we give in Table \ref{table:classbest:correction} the re-classification of these objects as a correction of the officially published catalogue. 

The distribution of the XID=1 objects in an X-ray/optical colour-colour diagram is shown in Fig.~\ref{f:stars_ccd}. The quantity plotted along the ordinate was computed as $\log(\texttt{RXS\_SRC\_FLUX}) + 0.4\times \texttt{SDSS\_MODELMAG\_i} + 5.61425$. The optical colour $g-r$ was built from the SDSS MODELMAG columns. The many objects in grey in the background are all identified objects in the catalogue (10404). The white dwarfs stick out as extreme blue objects with a high X-ray to optical flux ratio. Many of the single stars are likely coronal emitters in late-type stars and to be found as red objects with $g-r \simeq 1.5$. 

The compact binaries appear on top of the abundant AGN with a median $g-r\simeq 0.2$ and a median $\log (f_{\rm X} / f_{\rm opt}) \simeq 0.9$ but with a large dispersion in both quantities. 
Among the compact white dwarf binaries that are not CVs we find three objects that were previously classified as WDMS objects \citep[detached white dwarf main sequence objects][]{heller+09,rebassa+12} and one magnetic pre-cataclysmic binary \citep[a so-called LARP - low accretion rate polar, ][]{schwope+02}. The origin of their X-ray emission needs to be addressed separately, as well as the extreme X-ray emission of a few of the apparently normal stars around $g-r\sim 0.7, \log (f_{\rm X} / f_{\rm opt}) \sim 0.5$. Such a discussion, together with a more thorough presentation of the stellar content of the survey, is foreseen in a subsequent paper. 

\subsection{XMMSL2}
For the SPIDERS-XMMSL2 stellar sources, the emerging picture is slightly different. We find 19/2/6 objects in the XID=1/2/3 bins, a much higher fraction of XID=1 candidates as in 2RXS. 
Among the 37 objects with {\tt CLASS\_BEST==``STAR''} we re-classify two as Blazar (still XID=1, although not being a star, 4385-55752-614, 8172-57423-839), and one further, following the arguments given above, as likely cluster of galaxies, which thus becomes an XID=3 object. 
In this case, XMMSL2\,J113224.0+555745 (8170-57131-926), the BCG of the cluster lies even closer to the X-ray position than the M-star whose spectrum was taken. 
Other objects classified as XID=3 were F, K or M stars which appeared way too faint given the measured X-ray flux.

Among the XID=1 sources, we find 16 CVs and only three late-type coronal emitters (M5, M6). Interestingly, the majority (11 out of 19) XID=1 sources of the SPIDERS-XMMSL2 have a likelihood of any association ${\tt p\_any} > 0.5$. It confirms that having a reliable X-ray positional error is key to obtain accurate counterparts. To resolve ambiguities mentioned in this section, it would appear advisable to additionally visualise X-ray contours on the optical (or infrared) finding charts, instead of just using coordinates.

\section{AGN Spectral Properties}\label{sec:SpectralAnalysis}

A detailed discussion of the optical spectral properties of the SPIDERS sample is beyond the scope of this paper. 
We refer the reader to \citet{Coffey2019,Wolf2019} for an exploration of the detailed properties of SPIDERS type 1 AGN with sufficient signal-to-noise ratio in individual spectra. 
\citet{Wolf2019} investigated the markers of optical diversity of Type 1 AGN by deriving the principal components of optical and X-ray features for a sample of sources identified in SDSS-IV/SPIDERS and compiled by \citet{Coffey2019}. 
Making use of the large redshift and luminosity ranges probed by the SPIDERS sample, they could confirm that the broad $\rm H\beta$ line shape significantly evolves along the main sequence of broad line AGN (for a review see \citealt{Marziani18}). \citet{Wolf2019} report that the scaling of the FeII and the continuum emission strengths strongly depends on the sign of the asymmetry of $\rm H\beta$. The effect is discussed in the light of Broad Line Region outflows.

Instead, we present here a description of the general features of the sample. A benefit from having a large number of spectra is in stacking similar objects to increase the signal-to-noise ratio
per pixel and possibly unveil new features in the spectra \citep[e.g.][]{zhu2015}. In the following, we stack SPIDERS-DR16 spectra to create templates for generic usage, for example, exposure time calculation for spectroscopy, redshift fitting re-simulation, etc. The stacks are made available here\footnote{\url{http://www.mpe.mpg.de/XraySurveys/SPIDERS/}}. 

On average, the signal-to-noise ratio per pixel grows with the number of spectra stacked together as follows:
\begin{equation}
\mathrm{\log_{10}(S/N\; per\; pixel) = 0.45( 1 + \log_{10}(N\; spec \; per\; pixel) )}
.\end{equation}
The median signal-to-noise ratio per pixel in the observed spectra is $10^{0.45}=2.81$. By stacking 3000 (1000) spectra one reaches a signal-to-noise ratio on the order of 100 (60). 

\subsection{Spectral stacking method}
\label{sec:stacking_method}

First, we translated each observed spectrum to its rest-frame $\lambda_{RF}=\lambda/(1+z)$. 
Then we interpolated each spectrum and its uncertainties on a fixed wavelength grid in log$_{10}$ wavelength between 800$\AA$ and 11,000$\AA$ with a $\Delta \log_{10}(\lambda)=0.0001$ using \textsc{spectres} \citep{Carnall2017}. 
Finally, we took the median value of all fluxes in each pixels to obtain a stacked spectrum on this wavelength grid. 
We estimated the uncertainty on the median flux with a jackknife procedure. 
We note that to each spectrum, a normalisation (or a weight, e.g. a luminosity function completeness weight) can be applied, but this feature was not used here. 
This stacking procedure was previously applied in \citep{zhu2015,Raichoor17,Huang2019,Zhang2019} to stack spectra from star-forming galaxy. 
It is also used to stack the spectra of passives galaxies observed in the SPIDERS-CLUSTERS programme. 
These are presented in Clerc et al. (in prep). 
The accuracy on the redshift of AGN being lower than that of star-forming galaxies (with narrow lines), some information spanning the width of a few pixels is washed out in the stacks; the broad features remain. We chose a redshift bin with width 0.2 (or 0.5) and slide the redshift window by 0.1 to obtain a consistent evolution between the stacks. 
If more than 100 spectra were available in a bin, then we computed the stack. 

\begin{figure*}
\centering
\includegraphics[width=19cm]{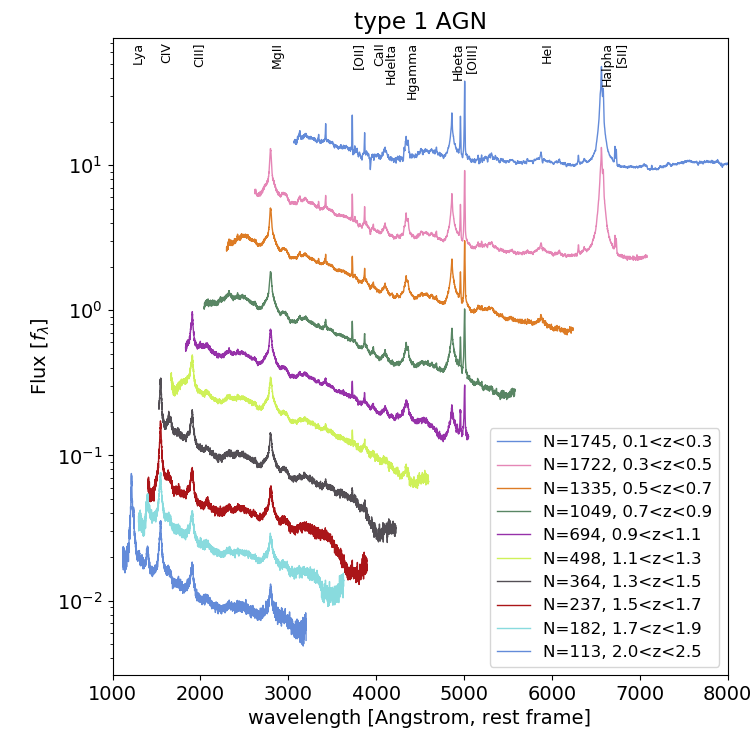}
\caption{Spectral stacks as a function of redshift for objects classified as type 1 AGN. Vertical displacement between spectra are added for clarity.}
\label{fig:spectral:stacks:qso:0}
\end{figure*}

\begin{figure*}
\centering
\includegraphics[width=19cm]{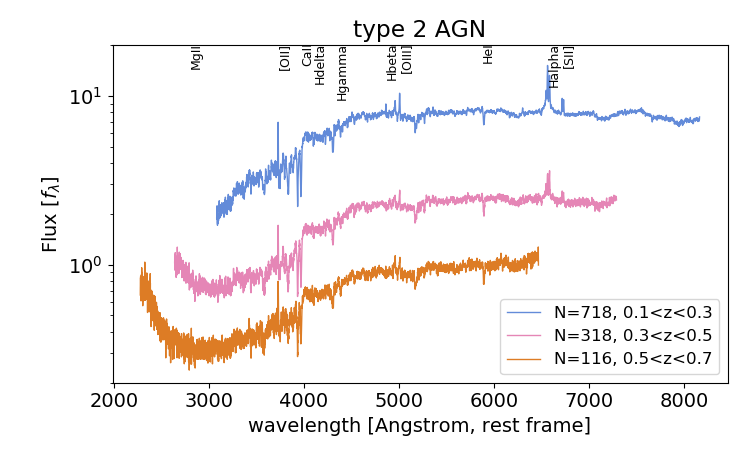}
\caption{Spectral stacks as a function of redshift for objects classified as type 2 AGN. Vertical displacement between spectra are added for clarity.}
\label{fig:spectral:stacks:type2}
\end{figure*}

\begin{figure*}
\centering
\includegraphics[width=19cm]{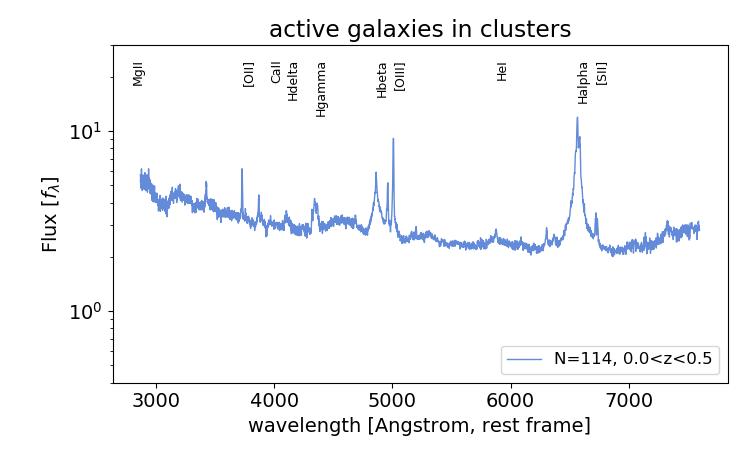}
\includegraphics[width=19cm]{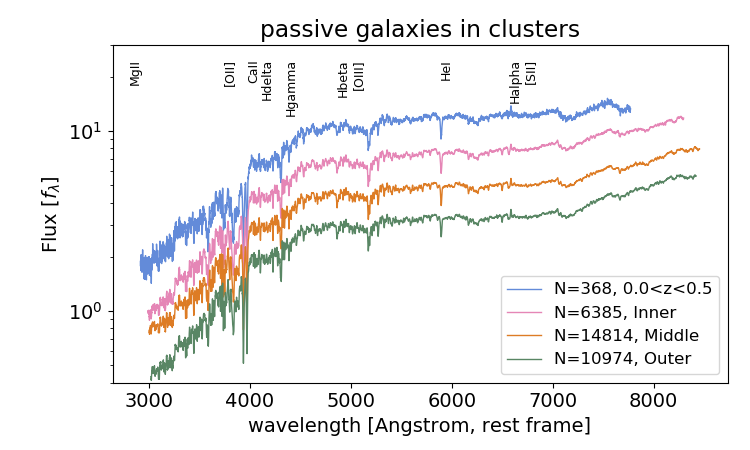}
\caption{Spectral stacks as a function of redshift for objects classified as galaxy in cluster. The top panel shows the stack of active galaxies in clusters and the bottom panel shows the stack of passive galaxies in clusters. The bottom panel is accompanied by the stack of passive galaxies in clusters from the SPIDERS-CLUSTER observations with their evolution as a function of cluster-centric radius, see Clerc et al. in preparation for full details.  Vertical displacement between spectra are added for clarity.}
\label{fig:spectral:stacks:clusterGalaxy}
\end{figure*}

\subsection{Type 1 AGN}
We selected type 1 AGN spectra in the 2RXS sample. 
There are enough spectra for the stacks to cover up to redshift 2.5. 
Figure \ref{fig:spectral:stacks:qso:0} shows the stacks obtained on a rest-frame wavelength axis in a $f_\lambda$ convention. 
The stacks obtained are consistent with the findings of \citet{VandenBerk2001}. 

We zoom in on the second and the last spectra to show the variety of features detected in Figs. \ref{fig:spectral:stacks:qso:zoom:a}, \ref{fig:spectral:stacks:qso:zoom:b}. 
We compare it to the SDSS DR5 spectral templates of the QSO (DR5 29) and of the luminous QSO (DR5 32) \citep{AdelmanMcCarthy2007}. Emission line features are more marked (higher equivalent widths) in the SPIDERS templates.

\subsection{Type 2 AGN}
In SPIDERS-DR16, the sample of type 2 AGN is large enough and spectroscopic data is homogeneous, so that we can create stacks up to redshift 0.7. 
We were previously lacking such stacks due to a smaller number of spectra or less homogeneous observations (exposure time, different instruments), which made the stacking procedure tedious. 
Fig. \ref{fig:spectral:stacks:type2} shows the stacks obtained. 
There, H$\alpha$ seems to be somewhat broad meaning that the type 2 classification is not perfect. 

\subsection{Galaxies in clusters}

Fig. \ref{fig:spectral:stacks:clusterGalaxy} shows the stacks of sources that are in the vicinity of optically detected clusters. 
The stacks show that we can separate (on average) the two populations of active galaxies in clusters and passive galaxies in clusters. 
The bottom panel is accompanied by the stack of passive galaxies in clusters from the SPIDERS-CLUSTER observations with their evolution as a function of cluster-centric radius, see Clerc et al. in preparation for full details. 
The stack of galaxies in clusters `contaminating' the AGN sample looks exactly like stacks of passive galaxies found to be cluster members. 

\subsection{Black hole mass and Eddington ratio}
The FWHM of $\rm H\beta$ frequently serves as a virial broadening estimator and is used to estimate black hole masses \citep[e.g.][]{Trakhtenbrot12, Mejia16}. 
The flux ratio $\rm r_{FeII}=F(FeII)/F(H\beta)$ is known to correlate with the Eddington ratio \citep{Grupe99,Marziani01,Du16}. 
These two parameters were initially among the main correlates of the original Eigenvector 1 (EV1), that is, the vector through optical and X-ray parameter space, which spans the most total variance \citep{Boroson92}. 
The plane $\rm FWHM_{H\beta}$ and $\rm r_{FeII}$ span is known as the EV 1 plane. 
The distribution of Type 1 AGN in this plane has been identified as main sequence of broad line AGN \citep[e.g.][and references therein]{Marziani18} and has proven of great use in the characterisation of the optical diversity of these sources. 
The stacking method described in this work can be applied in this context by using the binning of the Eigenvector 1 plane proposed by \citet{Sulentic02}. \citet{Sulentic02} as well as \citet{Zamfir10} have computed median composite spectra to investigate the evolution of the broad $\rm H\beta$ line shape along the EV1 sequence. 
The large number of sources available from the SPIDERS programme can be used similarly to uncover the dominating trends in the Balmer line diagnostics with increasing black hole mass and increasing Eddington ratio. 
In order to demonstrate the high S/N achieved with our stacks, we made use of the DR16 update of the SDSS-IV/SPIDERS Type 1 AGN catalogue compiled by \citet{Coffey2019}. 
$\rm FWHM_{H\beta}$ and $\rm r_{FeII}$ are listed as derived parameters in the catalogue from \citet{Coffey2019} and we identified sources in the following bins : 
\begin{itemize}
    \item A1: $\rm  0 \, km \, s^{-1}<FWHM_{H\beta} < 4000 \, km \, s^{-1}$ and $\rm 0 <r_{FeII}< 0.5$
    \item B1: $ \rm 4000 \, km \, s^{-1}  <FWHM_{H\beta} < 8000 \, km \, s^{-1}$ and $\rm 0<r_{FeII}<0.5$
    \item B1+: $\rm  80000 \, km \, s^{-1}<FWHM_{H\beta} <12000 \, km \, s^{-1}$ and $\rm 0<r_{FeII}<0.5$
\end{itemize}

\begin{figure*}
\centering
\includegraphics[width=2\columnwidth]{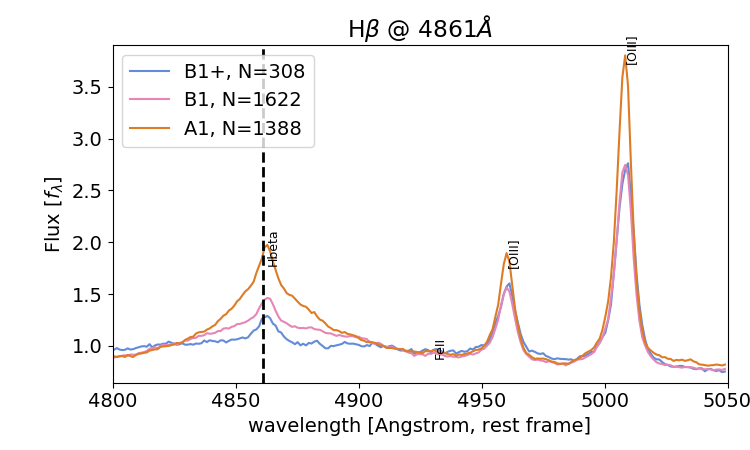}
\caption{Spectral stacks for objects classified as type 1 AGN. 
A zoom on the $\rm H\beta$ spectral region is presented.  
The stacks are divided by their median flux to ease comparison.
Stacks are taken in bins along the vertical EV1 sequence \citet{Sulentic02} with the emission line from \citet{VandenBerk2001} marked.}
\label{fig:spectral:stacks:qso:EV1}
\end{figure*}

The spectra of these sources were stacked following the method described in section \ref{sec:stacking_method}. Figure \ref{fig:spectral:stacks:qso:EV1} zooms on the $\rm H\beta$ line in these stacks. To guide the eye, we overplot the location of emission lines \citep{VandenBerk2001}. 
For increasing FWHM of $\rm H\beta$ one can clearly see the gradual appearance of a distinct very broad, slightly redshifted component in the stacked $\rm H\beta$, confirming the results by \citet{Sulentic02,Zamfir10}. 
Finer bins in the EV1 plane or further key optical parameter planes will allow us to probe the physics and geometry of the Broad Line Region in future work.

\section{Conclusions and outlook}\label{sec:Discussion}

In this work, we present the contents of the optical spectroscopic catalogue associated to X-ray point-like sources in the SPIDERS survey, published as part of the SDSS DR16. The systematic, highly complete follow-up programme assembled within four generations of SDSS delivers the largest spectroscopic redshift sample of an X-ray survey to date and represents a test-bed for a large programme of identification for large X-ray surveys in the future, especially with regard to the upcoming eROSITA all-sky survey.

The combination of wide-area X-ray surveys with optical spectroscopy enables a large number of unique scientific applications. As a further example, we a possible application for cosmology discuss below, following the works of \citet{Risaliti2015,Lusso2017}.

\subsection{Future AGN spectroscopic surveys following X-ray selected AGNs}
\label{subsec:future:surveys}

The SRG eROSITA full-sky scans will provide large number of targets for spectroscopic observation \citep{Merloni12,Predehl16}. 
SDSS-IV SPIDERS has demonstrated its ability to observed AGN with high completeness and to unambiguously classify the X-ray sources. 
For eROSITA, eRASS8 with a flux limit around -14, the peak of the number density should be around $z\sim1$ \citep{Merloni12,Comparat19} (compared to 0.1-0.2 in 2RXS, XMMSL2). 
It justifies the need of larger spectroscopic infrastructure to be complete. 

The next X-ray observation programme lined up is a transition programme linking SDSS-4 and SDSS-5. 
This programme is named eFEDS and will consist in 12 eROSITA dedicated plates covering 60 deg$^2$ within the footprint of the eROSITA Performance Verification programme. The data will be released as part of the next SDSS Data Release. 

Later in September 2020, following the completion of the first full-sky scan, SDSS-V \citet{Kollmeier17}, with its telescopes located in both hemisphere, will optimally observe the bright half of the sources. 
A couple of years later, using  a deeper four-year full-sky scan, 4MOST \citep{Merloni2019} will observe the fainter half of the eROSITA sources. 

In the longer term, the Athena \citep{Nandra2013athena} observatory will be well matched to the capabilities of the upcoming optical multi-fiber spectrograph MSE facility \citep{McConnachie16} to be mounted on a 10 meter class telescope. 

\subsection{A tentative forecast for the eROSITA era: cosmology with the AGN standard candle}

Recently, \citet{Risaliti2015,Lusso2017} proposed a method to construct quasar standard candles. 
It relies on the fact that exist an intrinsic non-linear relation between the UV emission from the accretion disk and the X-ray emission from the surrounding corona of the AGN. 
This relation between X-ray luminosity and UV luminosity has been observed \citep{Lusso2010}. 
Our current understanding of the disc-coronae and its non-linear scaling between UV and X-ray luminosity is not yet sufficient to prove this method in details. In the literature, there is skepticism about the physical disc-coronae model from \citet{Lusso2017} to account for this relation. 
For example, \citet{Kubota2018}; \citet{Panda2019} propose a model that is in agreement with the LX-LUV relation from Lusso. 
On the contrary, after exploring the physics of the disc and the coronae within a radiatively efficient AGN model, \citet{Arcodia2019} could not find a satisfactory explanation for the tight relation observed. 
Some authors find the $\alpha$ OX to be correlated with the Eddigton ratio \citep{Lusso2010}, some authors do not \citep[e.g.][]{Vasudevan2009}; and others find a correlation with black hole mass. So this point is yet to be entirely proven.

SDSS-IV SPIDERS has demonstrated our ability to observed AGN with high completeness and to unambiguously classify the X-ray sources, as required by this method. 
Additionally, \citet{Coffey2019} showed our ability to determine accurately the relevant spectral features for such an analysis. 
A cosmological analysis of the SPIDERS 2RXS sample is limited by depth the X-ray data, as the X-ray properties of the AGN are not determined well enough and impede the best selection of type 1 AGN standard candles. 
In the near future, eROSITA will provide the necessary high quality X-ray data, and we estimate below the possibility of a cosmological constraints via this method using the eROSITA mock catalogue produced by \citet{Comparat19}. 
For all type 1 mock AGN, we simulate the quasar UV\,-\,X-ray relationship and derive distance modulus estimates following the \citet{Risaliti2015} method.
The resulting quasar Hubble diagram is then fit using a standard $\Lambda$CDM cosmological model to place constraints on $\rm \Omega_{M}$ and $\rm \Omega_{\Lambda}$. 
We use only sources for which 4MOST will obtain optical spectra at a signal-to-noise greater than or equal to 10. 
Among these sources, we assume that $\sim$10 per cent of these sources will have reliable measurements of both the UV and X-ray flux densities (conservative assumption). We find a best-fit cosmology compatible with the input cosmology of the simulations. 
The uncertainty obtained is 5\% on $\rm \Omega_{M}$ and 10\% $\Omega_{\Lambda}$. 
For comparison, \citet{Risaliti2015} with current samples constrained $\rm \Omega_{M}$ and $\Omega_{\Lambda}$ to the $\sim$40\% and 27\% level while the Union 2.1 supernovae sample from \citep{Suzuki2012} constrained them to the 14 and 11\% level. 

Given the large number of type 1 AGN to be detected by eROSITA, which will then be observed with optical spectroscopy, the combination of eROSITA + SDSS-5 and 4MOST should be able to unveil if the method is correct. 
If the method is proven right, it would produce competitive and independent constraints on cosmological parameters. 
More accurate forecasts where one simulates jointly the photometry and the spectroscopy, based on the stacks presented here, to populate the Hubble diagram are foreseen in upcoming studies (PhD Thesis of D. Coffey, to be submitted). 

\section*{Acknowledgements}
We thank Jan Kurpas and Fabian Emmerich (AIP) for help with data presentation for screening and plotting. 
We thank the referee for the constructive feedback. 

This paper represents an effort by both the SDSS-IV collaborations.
Funding for SDSS-III was provided by the Alfred
P. Sloan Foundation, the Participating Institutions, the
National Science Foundation, and the U.S. Department
of Energy Office of Science.
Funding for the Sloan Digital Sky Survey IV has been provided by
the Alfred P. Sloan Foundation, the U.S. Department of Energy Office of
Science, and the Participating Institutions. SDSS-IV acknowledges
support and resources from the Center for High-Performance Computing at
the University of Utah. The SDSS web site is www.sdss.org.
SDSS-IV is managed by the Astrophysical Research Consortium for the
Participating Institutions of the SDSS Collaboration including the
Brazilian Participation Group, the Carnegie Institution for Science,
Carnegie Mellon University, the Chilean Participation Group,
the French Participation Group, Harvard-Smithsonian Center for Astrophysics,
Instituto de Astrof\'isica de Canarias, The Johns Hopkins University,
Kavli Institute for the Physics and Mathematics of the Universe (IPMU) /
University of Tokyo, Lawrence Berkeley National Laboratory,
Leibniz Institut f\"ur Astrophysik Potsdam (AIP),
Max-Planck-Institut f\"ur Astronomie (MPIA Heidelberg),
Max-Planck-Institut f\"ur Astrophysik (MPA Garching),
Max-Planck-Institut f\"ur Extraterrestrische Physik (MPE),
National Astronomical Observatory of China, New Mexico State University,
New York University, University of Notre Dame,
Observat\'ario Nacional / MCTI, The Ohio State University,
Pennsylvania State University, Shanghai Astronomical Observatory,
United Kingdom Participation Group,
Universidad Nacional Aut\'onoma de M\'exico, University of Arizona,
University of Colorado Boulder, University of Portsmouth,
University of Utah, University of Virginia, University of Washington,
University of Wisconsin,
Vanderbilt University, and Yale University.

This publication makes use of data products from the \textit{Wide-field Infrared Survey Explorer}, which is a joint project of the University of California, Los Angeles, and the Jet Propulsion Laboratory/California Institute of Technology, and NEOWISE, which is a project of the Jet Propulsion Laboratory/California Institute of Technology.
\textit{WISE} and NEOWISE are funded by the National Aeronautics and Space Administration.

The Pan-STARRS1 Surveys (PS1) and the PS1 public science archive have been made possible through contributions by the Institute for Astronomy, the University of Hawaii, the Pan-STARRS Project Office, the Max-Planck Society and its participating institutes, the Max Planck Institute for Astronomy, Heidelberg and the Max Planck Institute for Extraterrestrial Physics, Garching, The Johns Hopkins University, Durham University, the University of Edinburgh, the Queen's University Belfast, the Harvard-Smithsonian Center for Astrophysics, the Las Cumbres Observatory Global Telescope Network Incorporated, the National Central University of Taiwan, the Space Telescope Science Institute, the National Aeronautics and Space Administration under Grant No.~NNX08AR22G issued through the Planetary Science Division of the NASA Science Mission Directorate, the National Science Foundation Grant No. AST-1238877, the University of Maryland, Eotvos Lorand University (ELTE), the Los Alamos National Laboratory, and the Gordon and Betty Moore Foundation.

\bibliographystyle{aa}
\bibliography{references}

\listofobjects 

 
\appendix

\section{Tables and catalogues}\label{sec:Catalogs}

The XID values for sources classified as stars are given in Table \ref{table:xid}. 
The class correction for X-ray sources incorrectly classified as stars after are given in the table \ref{table:classbest:correction}.

\longtab[1]{
\begin{longtable}{ r r r  r r}
    \caption{\label{table:xid}
XID inspection flag for object correctly classified as STAR in the CLASS\_BEST classification. 
We use the following acronyms to classify the sources. 
CV: cataclysmic variable with unknown sub-category. 
CV/AM: cataclysmic variable of AM Herculis type. 
CV/DN: cataclysmic variable of dwarf nova type. 
WDMS: detached white dwarf/main sequence binary. 
LARP: low accretion rate polar. 
DB+M: a binary consisting of a white dwarf of spectral type DB and a companion star. }\\
\hline \hline
PLATE & MJD & FIBERID & XID & sub class \\ 
\hline 
\endfirsthead
\caption{Continued.} \\
PLATE & MJD & FIBERID & XID & sub class \\ 
\hline
\endhead
\hline
\endfoot
\hline
\endlastfoot
XMMSL2 \\
\hline
403 & 51871 & 423 & 1 & CV/AM  \\
444 & 51883 & 619 & 1 & CV/DN  \\
876 & 52669 & 103 & 1 & CV/DN  \\
973 & 52426 & 97 & 1 & CV/DN  \\
1172 & 52759 & 212 & 1 & CV/DN  \\
1997 & 53442 & 491 & 1 & WDMS   \\
2251 & 53557 & 606 & 2 & K1     \\
2623 & 54328 & 193 & 1 & CV     \\
2911 & 54631 & 598 & 2 & F9     \\
3176 & 54832 & 453 & 2 & K1     \\
3262 & 54884 & 508 & 1 & LARP   \\
5135 & 55862 & 59 & 1 & CV     \\
6445 & 56366 & 172 & 1 & CV/DN  \\
6687 & 56602 & 108 & 1 &        \\
6722 & 56431 & 820 & 1 & CV     \\
7624 & 57039 & 824 & 2 & F3     \\
7626 & 56934 & 516 & 2 & M1     \\
7650 & 57575 & 866 & 1 & CV/DN  \\
7677 & 57363 & 340 & 1 & M5     \\
7703 & 57333 & 818 & 1 & CV/AM  \\
7722 & 58425 & 504 & 1 & CV     \\
7734 & 58133 & 154 & 2 & M1     \\
7737 & 57722 & 164 & 1 & CV/AM  \\
7746 & 58074 & 680 & 2 & M5     \\
7891 & 57332 & 936 & 2 & K5     \\
8218 & 57519 & 860 & 3 & M1     \\
8748 & 58396 & 249 & 1 & M6     \\
9161 & 57691 & 615 & 1 & CV     \\
9174 & 58070 & 114 & 1 & CV     \\
10740 & 58223 & 592 & 1 & CV/AM  \\
10754 & 58224 & 381 & 2 & M3     \\
11053 & 58437 & 130 & 1 & M5     \\ 
\hline
2RXS \\
\hline
380 & 51792 & 575 & 1 & CV/DN  \\
384 & 51821 & 201 & 1 & M6     \\
384 & 51821 & 389 & 2 & M2     \\
403 & 51871 & 423 & 1 & CV/AM  \\
444 & 51883 & 619 & 1 & CV/DN  \\
542 & 51993 & 162 & 3 & G2     \\
550 & 51959 & 530 & 1 & DB+M   \\
551 & 51993 & 564 & 3 & F9     \\
619 & 52056 & 437 & 1 & LARP   \\
687 & 52518 & 262 & 1 & M6     \\
692 & 52201 & 47 & 1 & M5     \\
696 & 52209 & 338 & 1 & M4     \\
759 & 52254 & 331 & 3 & K7     \\
767 & 52252 & 262 & 2 & M3     \\
790 & 52441 & 82 & 3 & M*     \\
796 & 52401 & 640 & 1 & M4     \\
831 & 52294 & 309 & 3 & M0     \\
875 & 52354 & 61 & 3 & M2     \\
876 & 52669 & 103 & 1 & CV/DN  \\
888 & 52339 & 181 & 3 & K5     \\
906 & 52368 & 167 & 2 & M4     \\
936 & 52705 & 301 & 3 & G0     \\
941 & 52709 & 99 & 3 & K7     \\
956 & 52401 & 215 & 3 & M4     \\
959 & 52411 & 32 & 1 & M4     \\
970 & 52413 & 571 & 2 & F5     \\
972 & 52435 & 422 & 3 & K7     \\
974 & 52427 & 396 & 1 & WDMS   \\
1010 & 52649 & 313 & 1 & M0     \\
1083 & 52520 & 192 & 3 & G2     \\
1124 & 52914 & 588 & 2 & M3     \\
1172 & 52759 & 212 & 1 & CV/DN  \\
1215 & 52725 & 570 & 1 & K7     \\
1317 & 52765 & 338 & 2 & M3     \\
1337 & 52767 & 607 & 3 & K3     \\
1340 & 52781 & 198 & 3 & F5     \\
1341 & 52786 & 510 & 2 & M2     \\
1346 & 52822 & 220 & 3 & K5     \\
1381 & 53089 & 51 & 1 & WD     \\
1399 & 53172 & 410 & 1 & M3     \\
1419 & 53144 & 290 & 2 & F9     \\
1485 & 52992 & 545 & 3 & M4     \\
1668 & 53433 & 238 & 3 & K7     \\
1683 & 53436 & 615 & 1 & M0     \\
1948 & 53388 & 365 & 1 & M4     \\
1955 & 53442 & 479 & 1 & M2     \\
1955 & 53442 & 529 & 1 & CV/DN  \\
1986 & 53475 & 463 & 2 & K7     \\
1997 & 53442 & 491 & 1 & WDMS   \\
2000 & 53495 & 203 & 1 & M5     \\
2003 & 53442 & 32 & 2 & F9     \\
2018 & 53800 & 340 & 2 & M5     \\
2020 & 53431 & 144 & 2 & K3     \\
2031 & 53848 & 71 & 1 & M5     \\
2105 & 53472 & 22 & 3 & F9     \\
2181 & 53524 & 87 & 3 & F9     \\
2255 & 53565 & 202 & 1 & WD     \\
2313 & 53726 & 167 & 3 & F9     \\
2356 & 53786 & 382 & 1 & M3     \\
2387 & 53770 & 87 & 1 & K3     \\
2557 & 54178 & 335 & 1 & WD hot \\
2623 & 54328 & 193 & 1 & CV     \\
2974 & 54592 & 315 & 1 & K1     \\
3000 & 54843 & 379 & 1 & K1     \\
3000 & 54843 & 61 & 1 & M0     \\
3003 & 54845 & 603 & 1 & K1     \\
3003 & 54845 & 254 & 1 & K1     \\
3166 & 54830 & 489 & 3 & G2     \\
3237 & 54883 & 7 & 3 & G0     \\
3240 & 54883 & 207 & 1 & M3     \\
3240 & 54883 & 468 & 2 & M1     \\
3318 & 54951 & 114 & 3 & M1     \\
3406 & 54970 & 111 & 3 & F9     \\
3454 & 55003 & 211 & 3 & A0     \\
3459 & 55007 & 349 & 2 & F5     \\
3480 & 54999 & 264 & 3 & K1     \\
3677 & 55205 & 120 & 3 & K3     \\
3694 & 55209 & 356 & 1 & CV/AM  \\
3852 & 55243 & 530 & 2 & M6     \\
3855 & 55268 & 692 & 1 & M6     \\
4232 & 55447 & 158 & 3 & M1     \\
4987 & 55746 & 150 & 1 & WD     \\
6037 & 56106 & 210 & 1 & WD     \\
6056 & 56092 & 672 & 3 & K3     \\
6445 & 56366 & 172 & 1 & CV/DN  \\
6482 & 56358 & 989 & 3 & K5     \\
6587 & 56537 & 129 & 2 & M5     \\
6606 & 56596 & 372 & 3 & K5     \\
6649 & 56364 & 388 & 1 & CV/AM  \\
6670 & 56389 & 993 & 3 & M4     \\
6687 & 56602 & 108 & 1 & CV/DN  \\
6722 & 56431 & 820 & 1 & CV     \\
6723 & 56428 & 78 & 3 & K5     \\
6744 & 56399 & 430 & 3 & K5     \\
7279 & 57071 & 398 & 2 & M4     \\
7280 & 56709 & 868 & 2 & K*     \\
7281 & 57007 & 796 & 3 & F8     \\
7289 & 57039 & 184 & 1 & M5     \\
7296 & 57046 & 374 & 3 & K5     \\
7311 & 57038 & 192 & 1 & CV/DN  \\
7315 & 56685 & 944 & 3 & M1     \\
7332 & 56683 & 270 & 2 & K3     \\
7395 & 57131 & 28 & 3 & K5     \\
7413 & 56769 & 342 & 2 & M3     \\
7419 & 56811 & 183 & 3 & M4     \\
7577 & 56944 & 440 & 2 & M4     \\
7578 & 56956 & 83 & 1 & CV/DN  \\
7578 & 56956 & 264 & 1 & CV/AM  \\
7578 & 56956 & 759 & 2 & M5     \\
7583 & 56958 & 349 & 1 & G4     \\
7586 & 57186 & 73 & 1 & CV     \\
7589 & 56946 & 234 & 1 & M5     \\
7590 & 56944 & 674 & 2 & M5     \\
7600 & 56984 & 357 & 2 & M4     \\
7601 & 56959 & 950 & 1 & M4     \\
7606 & 56977 & 478 & 3 & M3     \\
7610 & 56980 & 894 & 2 & M5     \\
7612 & 56972 & 186 & 3 & M3     \\
7614 & 57307 & 844 & 3 & K3     \\
7619 & 56900 & 530 & 2 & M4     \\
7646 & 57570 & 238 & 2 & M1     \\
7647 & 57655 & 205 & 1 & K3     \\
7647 & 57655 & 853 & 2 & M3     \\
7659 & 57060 & 195 & 1 & M4     \\
7670 & 57328 & 618 & 3 & K5     \\
7681 & 57042 & 703 & 1 & M3     \\
7686 & 57015 & 427 & 3 & K5     \\
7688 & 57360 & 53 & 3 & M4     \\
7689 & 57743 & 453 & 3 & M1     \\
7693 & 57361 & 268 & 2 & K5     \\
7694 & 57359 & 325 & 1 & CV/DN  \\
7703 & 57333 & 818 & 1 & CV/AM  \\
7716 & 58097 & 923 & 2 & M1     \\
7722 & 58425 & 504 & 1 & CV     \\
7723 & 58430 & 757 & 1 & M5     \\
7723 & 58430 & 650 & 2 & M5     \\
7724 & 58434 & 232 & 1 & M5     \\
7725 & 58158 & 996 & 1 & M4     \\
7725 & 58158 & 725 & 1 & M5     \\
7728 & 58138 & 661 & 3 & G4     \\
7729 & 58135 & 239 & 1 & K0     \\
7731 & 58130 & 177 & 1 & CV/DN  \\
7732 & 58108 & 14 & 3 & G4     \\
7735 & 58136 & 329 & 1 & M4     \\
7735 & 58136 & 526 & 1 & CV/DN  \\
7736 & 57728 & 35 & 1 & M5     \\
7737 & 57722 & 164 & 1 & CV/AM  \\
7738 & 58100 & 351 & 1 & CV     \\
7748 & 58396 & 910 & 2 & M4     \\
7752 & 58072 & 990 & 1 & M5     \\
7757 & 58392 & 264 & 1 & M5     \\
7758 & 58402 & 635 & 2 & K5     \\
7759 & 58401 & 801 & 3 & M1     \\
7759 & 58401 & 644 & 2 & M3     \\
7765 & 58047 & 703 & 1 & CV     \\
7766 & 58395 & 683 & 3 & K0     \\
7767 & 58049 & 935 & 1 & CV     \\
7818 & 56989 & 152 & 3 & G4     \\
7845 & 56980 & 266 & 1 & K5     \\
7852 & 56987 & 252 & 2 & M4     \\
7854 & 56989 & 807 & 3 & G4     \\
7856 & 57260 & 350 & 1 & M5     \\
7860 & 57006 & 489 & 1 & M5     \\
7867 & 57003 & 570 & 2 & K3     \\
7869 & 57012 & 529 & 1 & M3     \\
7872 & 57279 & 849 & 1 & M4     \\
7879 & 57359 & 94 & 2 & M5     \\
7892 & 57333 & 823 & 2 & M1     \\
7912 & 57310 & 463 & 3 & G0     \\
7913 & 57333 & 432 & 1 & M5     \\
8054 & 57194 & 883 & 3 & K3     \\
8065 & 58248 & 642 & 3 & M5     \\
8068 & 57185 & 412 & 3 & K0     \\
8068 & 57185 & 743 & 3 & K5     \\
8160 & 57071 & 812 & 2 & M1     \\
8161 & 57127 & 768 & 3 & K5     \\
8177 & 57374 & 186 & 2 & M4     \\
8181 & 57073 & 201 & 3 & M*     \\
8181 & 57073 & 754 & 1 & M5     \\
8184 & 57426 & 636 & 3 & M3     \\
8187 & 57423 & 380 & 2 & M3     \\
8196 & 57346 & 569 & 2 & M5     \\
8276 & 57067 & 882 & 3 & M1     \\
8278 & 56990 & 182 & 3 & K5     \\
8281 & 57042 & 857 & 3 & M5     \\
8282 & 57041 & 565 & 3 & M4     \\
8287 & 57401 & 800 & 3 & K5     \\
8288 & 57419 & 680 & 3 & M5     \\
8297 & 57416 & 578 & 1 & WD     \\
8307 & 57723 & 162 & 3 & M4     \\
8308 & 57417 & 349 & 3 & M1     \\
8359 & 57449 & 474 & 2 & M4     \\
8376 & 57786 & 719 & 2 & M4     \\
8380 & 57520 & 710 & 3 & K5     \\
8406 & 57513 & 5 & 3 & K5     \\
8406 & 57513 & 261 & 3 & M1     \\
8408 & 57874 & 150 & 3 & G4     \\
8418 & 58199 & 453 & 2 & M*     \\
8425 & 58226 & 848 & 3 & K3     \\
8426 & 58224 & 773 & 1 & CV/DN  \\
8430 & 57488 & 46 & 3 & K5     \\
8491 & 57488 & 8 & 3 & WDMS   \\
8492 & 58171 & 650 & 3 & K5     \\
8500 & 57432 & 406 & 1 & M5     \\
8505 & 57834 & 239 & 2 & M1     \\
8515 & 58192 & 506 & 1 & M6     \\
8515 & 58192 & 850 & 3 & F8     \\
8517 & 57899 & 491 & 3 & K3     \\
8528 & 57896 & 668 & 3 & M1     \\
8528 & 57896 & 792 & 2 & M4     \\
8533 & 58017 & 920 & 1 & M5     \\
8536 & 58015 & 810 & 3 & K3     \\
8541 & 58257 & 330 & 3 & M3     \\
8735 & 58133 & 416 & 2 & M1     \\
8740 & 57367 & 520 & 3 & K0     \\
8823 & 57446 & 758 & 1 & M6     \\
8823 & 57446 & 408 & 1 & M5     \\
8832 & 57445 & 591 & 2 & M4     \\
8837 & 57867 & 899 & 2 & M3     \\
8848 & 57875 & 909 & 2 & M1     \\
8860 & 57458 & 922 & 1 & M4     \\
8868 & 57781 & 901 & 3 & M1     \\
8876 & 57783 & 90 & 2 & M1     \\
8877 & 57782 & 324 & 3 & M1     \\
8878 & 57785 & 614 & 3 & F8     \\
9146 & 58042 & 56 & 3 & K5     \\
9174 & 58070 & 492 & 2 & M4     \\
9178 & 58081 & 208 & 1 & M5     \\
9361 & 58055 & 740 & 2 & M1     \\
9363 & 57742 & 377 & 3 & M4     \\
9364 & 57699 & 922 & 3 & G8     \\
9395 & 58113 & 127 & 2 & F2     \\
9594 & 58135 & 194 & 1 & CV     \\
10241 & 58157 & 660 & 3 & F4     \\
10243 & 58159 & 870 & 1 & K5     \\
10250 & 58472 & 586 & 2 & K3     \\
10251 & 58173 & 313 & 3 & M5     \\
10265 & 58512 & 288 & 2 & M4     \\
10271 & 58497 & 304 & 1 & M5     \\
10289 & 58133 & 326 & 1 & M5     \\
10430 & 58155 & 271 & 1 & CV     \\
10723 & 58287 & 950 & 2 & M5     \\
10725 & 58250 & 892 & 2 & M3     \\
10726 & 58199 & 850 & 3 & M1     \\
10728 & 58248 & 597 & 2 & M5     \\
10740 & 58223 & 659 & 1 & M5     \\
10752 & 58488 & 14 & 1 & G8     \\
10901 & 58397 & 990 & 2 & M4     \\
10902 & 58396 & 414 & 2 & M4     \\
10908 & 58392 & 39 & 3 & F8     \\
10912 & 58253 & 510 & 1 & M1     \\
10913 & 58256 & 84 & 3 & K5     \\
10917 & 58252 & 909 & 2 & M4     \\
10917 & 58252 & 106 & 2 & M4     \\
11042 & 58462 & 144 & 3 & K3     \\
11076 & 58428 & 499 & 2 & M3     \\
11077 & 58433 & 817 & 1 & M5     \\
11086 & 58401 & 216 & 1 & M1     \\
11113 & 58425 & 610 & 2 & M3     \\
11125 & 58433 & 990 & 3 & K5     \\
11277 & 58450 & 785 & 3 & K3     \\
11277 & 58450 & 342 & 1 & M5     \\
11277 & 58450 & 62 & 1 & M6     \\
11304 & 58448 & 868 & 3 & M3     \\
11306 & 58450 & 802 & 2 & M1     \\
11311 & 58429 & 220 & 2 & G4     \\
11317 & 58398 & 186 & 1 & M5     \\
11340 & 58433 & 766 & 1 & M5     \\
11345 & 58428 & 865 & 1 & F8     \\
11345 & 58428 & 478 & 1 & G0     \\
11347 & 58440 & 333 & 2 & M5     \\
11382 & 58456 & 211 & 3 & K5     \\
11634 & 58484 & 726 & 2 & K5     \\
11675 & 58523 & 808 & 2 & M4     \\
\end{longtable}
}

\begin{table}
    \centering
    \caption{
Set of X-ray sources incorrectly assigned to a neighbouring STAR. 
The X-ray entry corresponding to the PLATE-MJD-FIBERID should be re-assigned to another spectrum that corresponds to the X-ray class given in this Table. 
The corrected X-ray classification is given in this Table.
The PLATE-MJD-FIBERID in this Table do correspond to stellar spectra, but it is unlikely that these stars have emitted the X-ray radiation. 
}
    \begin{tabular}{ l r r r r}
\hline \hline
PLATE & MJD & FIBERID & X-ray class \\ \hline 
8172 & 57423 & 839 & BLAZAR  \\
8170 & 57131 & 926 & CLUSTER \\
1046 & 52460 & 78  & CLUSTER \\
7905 & 57666 & 790 & QSO     \\
5191 & 56065 & 826 & QSO     \\
8172 & 57423 & 839 & BLAZAR  \\
8286 & 57062 & 832 & NONE    \\
693 & 52254 & 599  & CLUSTER \\
8436 & 57895 & 969 & NONE    \\
9430 & 58112 & 699 & BLAZAR  \\
7680 & 58131 & 247 & BLAZAR  \\
11354 & 58441 & 257 & BLAZAR  \\
\hline
\end{tabular}
\label{table:classbest:correction}
\end{table}

\begin{figure*}
\centering
\includegraphics[width=1.7\columnwidth]{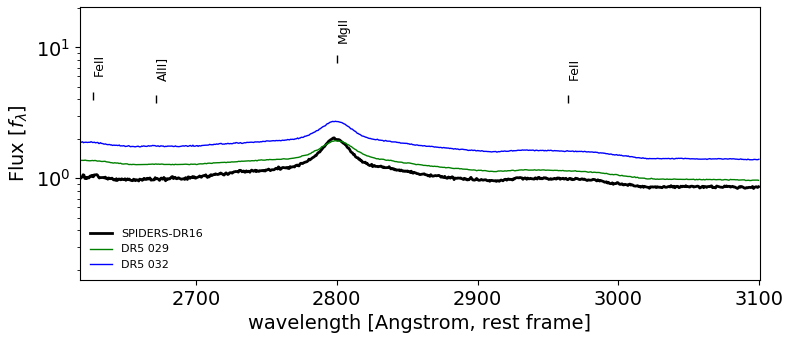}
\includegraphics[width=1.7\columnwidth]{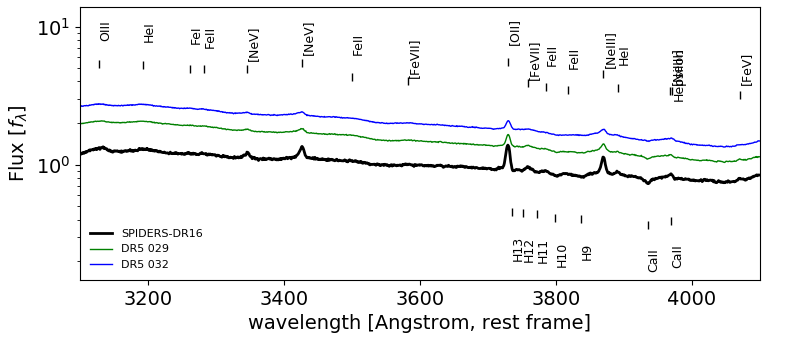}
\includegraphics[width=1.7\columnwidth]{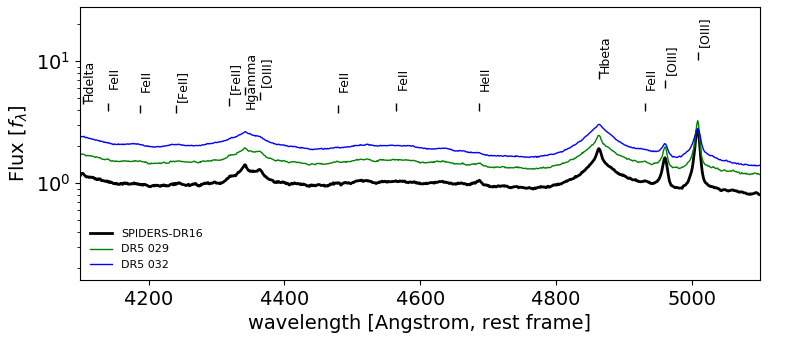}
\caption{Zoom on the wavelength range $2100<\lambda<5100$ of the type 1 AGN stack obtained over the redshift range $0.3<z<0.5$ with the emission line from \citet{VandenBerk2001} marked.
For comparison, we show the SDSS DR5 spectral templates of the QSO (DR5 29) and of the luminous QSO (DR5 32) \citep{AdelmanMcCarthy2007}. Emission line features are more marked (higher equivalent widths) in the SPIDERS templates. Vertical displacement between spectra are added for clarity. 
}
\label{fig:spectral:stacks:qso:zoom:a}
\end{figure*}
\begin{figure*}
\centering
\includegraphics[width=1.7\columnwidth]{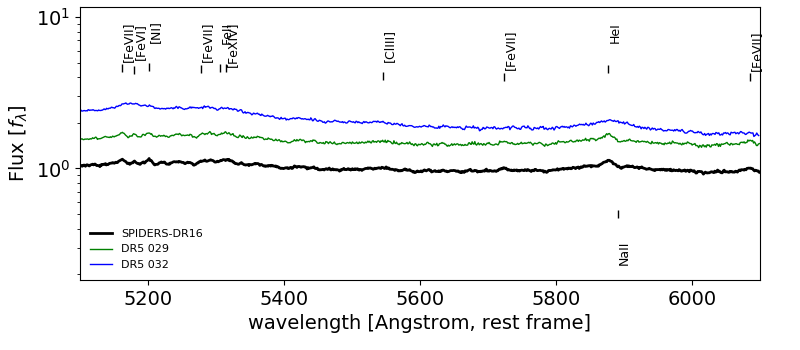}
\includegraphics[width=1.7\columnwidth]{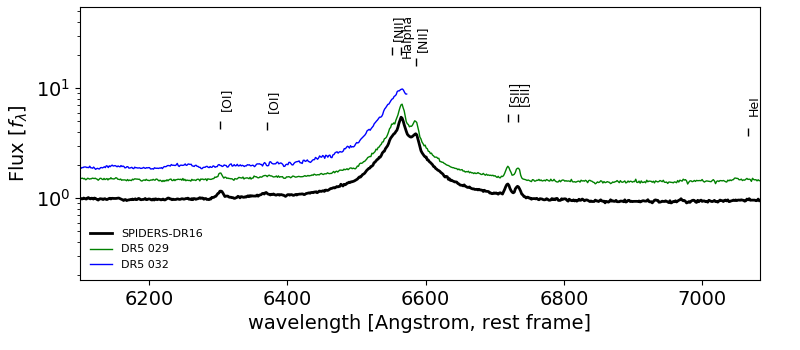}
\caption{Same as Figure \ref{fig:spectral:stacks:qso:zoom:a}. Zoom on the wavelength range $5100<\lambda<7100$ of the type 1 AGN stack obtained over the redshift range $0.3<z<0.5$. Vertical displacement between spectra are added for clarity. }
\label{fig:spectral:stacks:qso:zoom:b}
\end{figure*}
\end{document}